\documentclass[12pt]{article}
\usepackage{amsfonts, amsmath,graphicx, amsthm, amssymb, cite, stmaryrd}
\numberwithin{equation}{section}
\newtheorem{proposition}{Proposition}
\numberwithin{proposition}{section}
\newtheorem{thm}{Theorem}
\numberwithin{thm}{section}

\numberwithin{definition}{section}
\newtheorem{rem}{Remark}
\numberwithin{rem}{section}
\newtheorem{lem}{Lemma}
\numberwithin{lem}{section}
\newcommand{\RR}{\mathrm{I\!R\!}}
\usepackage[margin=1in]{geometry}
\usepackage{color}
\begin{document}
\title{On the Structure of Linear Dislocation Field  Theory}
\author{ A Acharya\thanks{Department of Civil and Environmental Engineering and Center for Nonlinear Analysis, Carnegie-Mellon University, Pittsburgh, PA 15213, USA, acharyaamit@cmu.edu.} \and R J Knops\thanks{The Maxwell Institute of Mathematical Sciences and School of Mathematical and Computing Sciences, Heriot-Watt University, Edinburgh, EH14 4AS, Scotland, UK, r.j.knops@hw.ac.uk.} \and J Sivaloganathan\thanks{Department of Mathematical Sciences, University of Bath, Bath, BA2 7AY, UK, J.Sivaloganathan@bath.ac.uk.} }

\date{}
\maketitle

\begin{abstract}
\noindent Uniqueness of solutions in the linear theory of non-singular dislocations, studied as a special case of plasticity theory, is examined. The status of the classical, singular Volterra  dislocation problem as a limit of   plasticity problems  is illustrated by
a specific example   that clarifies
the use of the plasticity formulation in the   study of   classical dislocation theory. Stationary, quasi-static, and dynamical problems for  continuous dislocation distributions  are investigated subject  not  only to  standard boundary and initial conditions, but also to prescribed dislocation density. In particular, the dislocation density field can represent a single dislocation line.

It is  only in the static and quasi-static traction boundary value problems that such data  are  sufficient for the  unique determination of   stress.  In other quasi-static boundary value problems and problems involving moving dislocations, the   plastic and elastic distortion tensors, total displacement, and stress  are in general non-unique
for specified dislocation density.  The conclusions are confirmed by the example of  a single screw  dislocation.
\end{abstract}



\section{Introduction}\label{intro}
 Dislocations in crystals are microstructural line defects  that  create `internal' stress even in the absence of  loads.  Physical observation suggests that applied loads and mutual interaction cause dislocations to move and   the body to  become permanently deformed. The understanding and prediction of the internal stress field and accompanying permanent deformation due to large arrays of dislocations form part of the fundamental study of metal plasticity.

In an elastic body, a dislocation is defined in terms of the non-zero line integral of the elastic distortion around possibly time-dependent closed curves or circuits in the body.
The elastic distortion itself  is related to the stress through a linear constitutive  assumption, while equilibrium requires the divergence of the stress to vanish in the absence of body-forces.   When the dislocations are continuously distributed, Kr\"{o}ner \cite{k81} uses Stokes' theorem
to derive the pointwise connection between the \textit{Curl}  operator of the elastic distortion and the dislocation density field.  The connection is  valid also for a single dislocation whether singular or not. Consequently, the elastic distortion is incompatible (i.e., is not the gradient of a vector field) in the theory of  continuously  distributed static  dislocations, which contrasts  with classical linear elasticity.
Furthermore, it must be shown how the elastic distortion can be determined from the equilibrium equations and elastic incompatibility for given dislocation density and linear elastic response.
Kr\"{o}ner adopts the dislocation density as data, but his proposed resolution of the problem determines only the symmetric part (the strain) of the elastic distortion.
 Willis \cite{willis1967second} observes  that the  boundary value problem with dislocation density as  data  in fact   determines the complete elastic distortion (including rotation). Kr\"oner also introduces  a total (generally continuous) displacement field  and defines the \emph{plastic distortion} as  the difference between the gradient of the displacement field  and the elastic distortion. 
The  approach is  motivated by \textit{cut-and-weld} operations \cite{nabarro1967theory, eshelby1956continuum, eshelby1957determination} used  to describe dislocations.

 Henceforth in this paper,  the  \textit{plasticity formulation, or theory,  (of dislocations)}  involves the total displacement,  stress,  plastic distortion, and dislocation density fields. Stress is  defined by  linear dependence on the elastic distortion subject to  the static or dynamic balance of forces.  The complete representation of  plasticity  due to moving
dislocations involves an  evolution equation for the plastic distortion. This  depends upon the stress state through a fundamental kinematical relation between the plastic distortion rate and  both  the dislocation density  and  stress-dependent velocity.

Thus, the stress and plastic distortion are intimately coupled. In this work, however, we simply determine the  stress and displacement  subject to    data that  includes  various parts of the   evolving plastic distortion field.  It is of particular interest to investigate  whether stress and displacement are uniquely determined when only the evolving  dislocation density field is  known. The topic is first encountered in  the equilibrium traction boundary-value problem
for which, as discussed later, 
the dislocation density  is sufficient  to uniquely determine the stress.

The classical  theory of dislocations,  developed  in   papers   \cite{michell1899direct, michell1899transmission, timpe1905probleme, weingarten1901sulle},  is due to Volterra \cite{volterra1907equilibre} and  does not deal with evolution. It  regards a dislocation    as the termination edge of  a surface over which the total displacement is discontinuous by an amount that defines the Burgers vector. The  traction 
remains  continuous across the surface. (See also \cite{love1944treatise,nabarro1967theory, hl82}.) The classical theory is stated in multiply-connected regions excluding dislocation cores. In such a  region, the displacement field of a dislocation may be viewed as a continuous multivalued `function.' Alternatively, and more conveniently, it may be viewed as a discontinuous function with  constant discontinuity across any surface whose removal   from the (multiply-connected) region renders the latter simply-connected. On the simply-connected region obtained by the use of such `barriers' or `cuts', the displacement may be regarded as  a single-valued  continuous vector field, which nevertheless has different values at adjacent points on either side of each  barrier.

The Volterra formulation contains singularities  not necessarily present in the plasticity formulation.  One task  therefore is  to reconcile the classical  and plasticity formulations. As a  first step, we explain  how 
a single stationary Volterra dislocation line is the formal limit 
of a sequence of problems in plasticity theory. Plasticity theory 
is a physically more realistic non-singular  description of moving dislocations and their fields.
It avoids mathematical difficulties  caused by nonlinearities in  non-integrable fields that would otherwise appear  in  the full problem of evolution coupled to stress.

 Apart from exploring  the relevance of the plasticity formulation to an understanding of dislocations, whether according to Kr\"oner's  or Volterra's interpretation, another major consideration  of this paper concerns  uniqueness  in the static and dynamic problems of  the  plasticity dislocation theory. Standard Cauchy initial conditions together with displacement, mixed, or traction boundary conditions are augmented  by a prescribed  dislocation density  rather than the usual plastic distortion tensor. 
A  previous contribution  \cite{willis1967second} demonstrates   that the stress     and elastic distorsion to within a constant skew-symmetric tensor  in the equilibrium traction boundary value problem on unbounded regions  are  unique subject  to a prescribed dislocation density field. It is noteworthy that this approach  dispenses entirely with the total displacement field.  In contrast,   it follows from  Weingarten's theorem  that the displacement  is  not unique  in   the classical  Volterra  theory  for a given dislocation distribution.  Uniqueness, however, can be retrieved   when the ``seat of the dislocation'' [Lov44], (the surface of displacement discontinuity) is additionally prescribed.

Time-dependent problems of plasticity in  a body containing a possibly large number of moving dislocation lines are physically important. They become prohibitively complicated when considering an excessively large number  of dislocations and their corresponding surfaces of discontinuity.
It then becomes convenient to replace arrays  of discrete dislocations by  continuous distributions of dislocations. An immediate difficulty, however, is encountered.
It is shown in \cite{a01,a03} that a prescribed dislocation density is insufficient to ensure  well-posedness of  the corresponding quasi-static traction boundary problem.
Elements  of the additional data necessary for well-posedness were subsequently simplified in \cite{roy2005finite} using  a decomposition of the elastic distortion similar to that of Stokes-Helmholtz. A   preliminary investigation in \cite[Sec.6c]{a03} and \cite[Sec.4.1.2]{roy2006size}  shows how,  when  deformation evolves, the stress is uniquely determined by the dislocation density in the corresponding quasi-static  traction boundary value  problem. The dislocation density, however, may no longer be sufficient to uniquely determine the stress in the quasi-static displacement or mixed  boundary value problems. 

A detailed analysis of uniqueness of solutions in sufficiently smooth function classes and the derivation of  new results are also among the aims of  this paper. Specifically, we separately treat the equilibrium dislocation traction boundary value problem, quasi-static boundary value problems, and  exact initial boundary value problems in which material inertia is retained. With inertia, a notable conclusion is that an evolving dislocation density field is insufficient to uniquely determine the stress in the initial boundary value problem subject to zero body force and zero boundary traction on regions that are bounded or unbounded. The result differs significantly from
the equilibrium and quasi-static traction boundary value problems where a prescribed dislocation density field is sufficient for uniqueness.
Extra conditions are derived for uniqueness  in those problems where a prescribed dislocation density is insufficient for the stress and displacement to be unique.

In this respect, the result of \cite{ma63, kosevich1979crystal} is accommodated in our approach. Our considerations  delineate the  deviations possible from the Mura-Kosevich proposal  provided  the evolving dislocation density  remains identical to theirs. Our treatment also enables a conventional problem in the phenomenological theory of plasticity to be interpreted in the context of dislocation mechanics.  We also  demonstrate that certain parts of two plastic distortions must be identical in order that the corresponding initial boundary value problems  possess identical  stress and displacement fields.

A subsidiary  task  is to explore  conditions for  the plasticity  dislocation theory to reduce to the respective classical linear elastic theories when the dislocation density vanishes. Included in the  necessary and sufficient conditions is the condition that the elastic distortion tensor field is the gradient of the classical displacement. For reasons explained later, we seek alternative sets of conditions which are  described in Sections~\ref{stat},\ref{qsbvp}, and \ref{mov}.

Various mathematical aspects of moving dislocations have been
developed and studied in \cite{e53, mb63, kosevich1979crystal, weertman1967uniformly,                                                                      nabarro1951cxxii, stroh1962steady, lazar2009gauge, lazar2013non, p10, lazar2016distributional, rosakis2001supersonic, m, markenscoff1990singular, ni2008self, willis1965dislocations, freund1998dynamic, clifton1981elastic, zhang2015single}, but none within the context  proposed here.  Of these contributions, those of Lazar are of closest interest. They suggest that besides an evolving dislocation density,  other elements are necessary to  satisfactorily  formulate theories of plastic evolution. 
Lazar applies the principle of  gauge invariance to  the underlying Hamiltonian of elasticity  theory. However, for  small deformations, the stress depends upon the  linearised rotation field \cite{lazar2009higgs}, and therefore  violates  invariance under rigid body deformation.  Parts of the discussions of Pellegrini and Markenscoff [Pel10,Pel11, Mar11] appear to be related to implications of our paper.

General notation, introductory concepts from dislocation theory, and  some other basic assumptions  are presented in   Section~\ref{notation}. 
Section~\ref{plast-Volterra-Love} discusses in detail an explicit example of a stationary straight Volterra dislocation and its relation to  plasticity theory. The example chosen consists of   a sequence of plastic distortion fields  defined on transition strips of vanishingly small width. Section~\ref{stat} considers the equilibrium traction boundary value problem for stationary dislocations, and confirms that the stress is unique for a prescribed dislocation density.
 As illustration,  a single static screw dislocation in the whole space is treated  by means of the Stokes-Helmholtz representation. A similar analysis to Section~\ref{stat}
is undertaken  in Section~\ref{qsbvp} for  the quasi-static problem with moving dislocations.  Now, however, for given dislocation density, the stress is unique  only in the traction boundary value problem. 
The total displacement and plastic distortion are non-unique for the traction, mixed, and displacement boundary value problems. Uniqueness of all three quantities (stress, elastic distortion and total displacement) is recovered when the plastic distortion is suitably restricted. Section~\ref{mov} derives separate  necessary and sufficient conditions for uniqueness of the stress and the total displacement fields  in the initial boundary value problem for moving dislocations with material inertia.  Section~\ref{mov}  further  identifies admissible   initial conditions for which the problem is physically independent of any special choice of reference configuration. In Section~\ref{scdisl} we discuss particular  initial value problems for  the single  screw dislocation uniformly moving in the whole space subject to specific, but natural, initial conditions. Explicit solutions 
demonstrate how the stress field may be non-unique for   prescribed evolving dislocation density and the same initial conditions.  Brief remarks in Section~\ref{conrem} conclude the paper.

\section{Notation and other preliminaries}\label{notation}

We adopt the standard conventions of a comma subscript to denote partial differentiation, and  repeated subscripts to indicate summation. Latin suffixes range over $1,2,3$, while Greek suffixes range over $1,2$, with the exception of the index $\eta$ which along with $t$ is reserved for the time variable.

Vectors and tensors are distinguished typographically by lower and upper case letters respectively,  except that $N$  is used to denote the unit outward vector normal on a surface.  A superscript  $T$  indicates  transposition,  while $M^{n\times m}$ denotes the set of real   $n\times m$ matrices. A direct and suffix notation is employed indiscriminately to represent vector and tensor quantities, with reliance upon the context for  precise meaning. Scalar quantities are not distinguished.  The symbol $\times$ indicates the cross-product, and a dot denotes the inner product.  Both symbols are variously used    for  products between vectors, vectors and tensors, and between tensors. 
The operator $grad$ applied to a scalar, and the operators $Grad,\,div,\, curl,$ and $Div,\,Curl,\,$  applied to vectors and second order tensors have their usual meanings. 
To be definite, with respect to a  common Cartesian rectangular coordinate system whose unit coordinate vectors form the set $(e_{1},\,e_{2},\,e_{3})$,  we have the formulae
\begin{align}
(A.N)_{i}& =A_{ij}N_{j} & &\\\label{tvprod}
 (u\times v)_{i} &=  e_{ijk}u_{j}v_{k}, &  (u\otimes v)_{ij} &= u_{i}v_{j},\\
 (grad\, \phi)_{i}&= \phi_{,i}, & (curl\, u)_{i}&= e_{ijk}u_{k,j},\\
div\, u&=u_{i,i},&  (A\times v)_{im} &= e_{mjk}A_{ij}v_{k},\\
 (A\times B)_{i} &= e_{ijk}A_{jr}B_{rk},& (Grad\,u)_{ij} & = (\nabla\,u)_{ij}=u_{i,j},\\
  (Div\,A)_{i} &=(\nabla . A)_{i} = A_{ij,j},&     (Curl\,A)_{im} & = (\nabla \times A)_{im} =e_{mjk}A_{ik,j},
\end{align}
where $e_{ijk}$ denotes the alternating tensor.

In addition, we require the following generalised functions and their derivatives (cp \cite{bracewell1978fourier}, pp 72-76). The Dirac delta function,  denoted by $\delta(x)$, possesses the properties
\begin{eqnarray}
\label{deltwo}
\delta(-x)&=&\delta(x),\\
\label{delthree}
f(x)\delta(x)&=& f(0)\delta(x),
\end{eqnarray}

where the function $f(x)$ is infinitely differentiable at $x=0$. Other generalised functions are the Heaviside  step funcion $H(x)$ and the sign function $sign(x)$   defined by
\begin{equation}\label{step}
\begin{split}
& H(x)=\begin{cases}
&      1 \qquad \mbox{when }\, x>0,\\
&      0\qquad \mbox{when }\, x \le 0,\\
      \end{cases}\\
\end{split}
\end{equation}
\begin{equation}\label{sign}
\begin{split}
&sign(x) = \begin{cases}
&          \ \ 1\qquad \mbox{when } \,x>0,\\
&         -1\qquad \mbox{when }\,x \le 0,\\
           \end{cases}\\
\end{split}
\end{equation}
 and which are related by
\begin{equation}
\label{hemstep}
sign(x)=2H(x)-1.
\end{equation}
 These generalised functions possess distributional derivatives,  indicated by a superposed prime, that satisfy
\begin{eqnarray}
\label{delH}
\delta(x)&=& H^{\prime}(x),\\
\label{sigdel}
sign^{\prime}(x)&=& 2\delta(x).
\end{eqnarray}

We also employ $A^s$ and $A^a$ to represent the symmetric and skew-symmetric parts, respectively, of the tensor $A$ so that
\begin{eqnarray}
A^{s}&=&\frac{1}{2}\left(A+A^{T}\right),\\
A^{a}&=& \frac{1}{2}\left(A-A^{T}\right).
\end{eqnarray}

Consider a region $\Omega\subseteq \RR^{n},\,n=2,3$ which may be unbounded or when bounded possesses the smooth boundary $\partial\Omega$ with unit outward vector   normal $N$. Unless otherwise stated, $\Omega$ is  simply connected  and  contains the  origin of the Cartesian coordinate system.

The region $\Omega$ is occupied by a (classical)  nonhomogeneous  anisotropic compressible linear elastic material whose elastic modulus tensor $C$ is differentiable and  possesses both major and minor symmetry so that the corresponding  Cartesian components satisfy
\begin{equation}
\label{majmin}
C_{ijkl}=C_{jikl}  =C_{klij},
\end{equation}
which imply the additional symmetry $C_{ijkl}=C_{ijlk}$. It is further supposed that the tensor $C$ is uniformly positive-definite in the sense that 
\begin{equation}
\label{conv}
c_{1}\phi_{ij}\phi_{ij}\le C_{ijkl}\phi_{ij}\phi_{kl},\qquad \forall\, \phi_{ij}=\phi_{ji}\neq 0. \qquad x\in\bar{\Omega},
\end{equation}

for positive constant $c_{1}$. In the illustrative examples, the elastic moduli are assumed constant for convenience.
The elastic body, subject to  zero applied body-force,  is   self-stressed  due to  an array of discrete dislocations  represented  by a continuous distribution of dislocations of  prescribed  density  denoted by the second order  tensor field $\alpha$. In   the stationary problem, the dislocation density is a spatially dependent continuously differentiable  tensor function.  For  time-dependent  problems, the density depends upon both space and time   so that  $\alpha(x,t)$ where $(x,t) \in\Omega \times [0,T)$ and  $[0,T)$ is the maximal interval of existence.

The prescription of the stationary problem is now considered in  detail.

 Let $\Sigma\subset \Omega$ be any open simple parametric surface bounded by the simple closed curve $\partial\Sigma$ described in a right-handed sense. The Burgers vector $b_{\Sigma}$ corresponding to the   patch $\Sigma$ is given by \cite{nye1953some, mb63,k81}
\begin{equation}
\label{bdis}
b_\Sigma=\int_{\Sigma}\alpha.dS,
\end{equation}
where $dS$ denotes the surface area element. The sign convention is  opposite to that adopted by most authors.
We introduce the second order non-symmetric \textit{elastic distortion} tensor  $U^{(E)}\in C^{1}(\Omega)$, as a second state variable. Its relation to Burgers vector is given by (c.p.,  \cite{k81}) 
\begin{eqnarray}
\label{beldis}
b_\Sigma &=& \oint_{\partial\Sigma} U^{(E)}.ds\\
\label{bsurel}
&=& \int_{\Sigma}Curl\, U^{(E)}.dS,
\end{eqnarray}
where Stokes' theorem is employed, and $ds$ denotes the curvilinear line element of $\partial\Sigma$.  Elimination of $b_\Sigma$ between (\ref{bdis}) and (\ref{bsurel}),  using the arbitrariness of $\Sigma$ , yields the fundamental field equation
\begin{equation}
\label{denel}
\alpha= Curl\, U^{(E)},\qquad x\in\Omega,
\end{equation}
from which is deduced the condition
\begin{equation}
\label{div}
Div\,\alpha =0,\qquad x\in\Omega.
\end{equation}

For non-vanishing dislocation density $\alpha$, (\ref{denel})  implies that $U^{(E)}$ is  incompatible  in the sense that there does not exist a twice continuously differentiable  vector field  $z(x)$ such that $U^{(E)}= Grad\, z,\,x\in\Omega$.     The relation  \eqref{denel} also implies that $\alpha$ determines $U^{(E)}$  only to within  the gradient of an arbitrary differentiable vector field. Determination of the components of $U^{(E)}$ that are uniquely specified by $\alpha$ forms  an essential  part of our  investigation.

The elastic distortion produces a stress distribution $\sigma(x)$ which according to Hooke's law and the symmetries (\ref{majmin}) is given by
\begin{equation}
\label{ssrel}
\sigma(x)=CU^{(E)}= C \left(U^{(E)}\right)^{s},\qquad x\in\Omega.
\end{equation}


Under zero body-force, the  stress $\sigma(x)$ in equilibrium satisfies  the  equations
\begin{equation}
\label{preeqeqn}
Div\,\sigma=0,\qquad x\in\Omega.
\end{equation}

Appropriate boundary conditions for the complete description of the stationary problem are  postponed to Section~\ref{stat}.

 We next discuss the  \textit{plastic distortion tensor} and consider  certain properties common to  both the  stationary and dynamic problems. 
Based upon a qualitative discussion of the formation of dislocations in crystals, Kr\"{o}ner \cite[\S 3]{k81}  defined the non-symmetric plastic distortion tensor  $U^{(P)}:\Omega \rightarrow M^{3\times 3}$ by the relation
\begin{equation}
\label{elplrelat}
U^{(P)}= Grad\, u-U^{(E)},
\end{equation}
where the vector field $u(x)$, assumed twice continuously differentiable in $\Omega$, is the \textit{total displacement}. Microcracks and similar phenomena are excluded  from consideration. The displacement field  $u(x)$ is  compatible and is produced by both external loads and dislocations. It is to be expected, but requires proof, that in the absence of dislocations,  $u(x)$ becomes the displacement field  of the classical linear theory, while in the absence of both dislocations and external loads, $u$ is identically zero. The topic is discussed for the stationary and dynamic problems in Sections~\ref{stat},~\ref{qsbvp}, and~\ref{mov}  where necessary and sufficient   conditions are  derived for the dislocation density $\alpha$ to vanish. One such set of conditions is simply $U^{(P)}=0$, but since the plastic distortion tensor is a postulated  state variable, we prefer to derive alternative necessary and sufficient conditions.

The elastic distortion may be eliminated between \eqref{elplrelat} and \eqref{denel} to obtain
\begin{eqnarray}
\nonumber
\alpha &=&Curl\,(Grad\,u-U^{(P)})\\
\label{ddpldistprel}
&=& -Curl\,U^{(P)}.
\end{eqnarray}
The plastic distortion tensor, incompatible when dislocations are present, is  often designated as data.  Our
objective, however,  is to  examine implications for uniqueness  when  the dislocation density is adopted as data and not the plastic distortion tensor. One immediate  difficulty apparent from \eqref{ddpldistprel} is that  the gradient of an arbitrary vector field may be added to $U^{(P)}$ without disturbing the equation.

Similar comments apply to the   initial boundary value problem containing the material inertia.  In this problem, the equations of motion for time-dependent stress $\sigma(x,t)$ subject to zero body force become
\begin{equation}
\label{preeqmtn}
Div\,\sigma =\rho\ddot{u},\qquad (x,t)\in\Omega\times [0,T),
\end{equation}
where $\rho$ denotes the mass density, and a superposed dot indicates time differentiation. 
Initial and boundary conditions for the dynamic problem are stated  in 
Section~\ref{mov}.

We recall  that a necessary and sufficient condition for the vanishing of the  strain tensor $e(u)$ given by
 \begin{equation}
 \label{rigstr}
 e(u)=\left(Grad\,u\right)^{s},\qquad e_{ij}(u)=\frac{1}{2}\left(u_{i,j}+u_{j,i}\right),
 \end{equation} 
 is that $u(x,t)$ is an infinitesimal rigid body motion, specified by
\begin{equation}
\label{rig}
u =a+x\times \widehat{\omega},
\end{equation}
where  $a(t),\,\widehat{\omega}(t)$ are vector functions of time alone. 

Sufficiency is obvious by direct substitution of (\ref{rig}) in (\ref{rigstr}). To prove neccessity, we define
\[
\omega(u) = (Grad u)^a, \qquad \omega_{ij}(u) =   \frac{1}{2} ( u_{i,j} - u_{j,i} ),
\]
and   note that for any twice-continuously differentiable vector field $u$ on $\Omega$ we have
\begin{equation}\label{korn_proxy}
2 \, \omega_{ik,l} = u_{i,kl} + u_{l,ki} - u_{l,ki} - u_{k,il} = 2 \, ( e_{il,k} - e_{kl,i}).
\end{equation}

Thus,  the rotation field of a displacement field is determined  by integration from  its strain field.  When $e(u) \equiv 0$, $\omega$ is at most a  time-dependent,  spatially constant skew-symmetric tensor function on $\Omega$.  The desired result (\ref{rig}) is obtained  by one spatial integration of the identity $Grad\,u=\omega$  and by letting $\widehat{\omega}$ be the axial vector of $\omega$.  Observe that (\ref{korn_proxy}) is the classical analog of Korn's inequality which states that the $H^1-$ norm of a vector field is bounded by a constant times the sum of the squares of the $L^2$ norms of the vector field and its strain field.  Accordingly, as just stated,  the rotation field of a displacement is controlled by its strain field.

We repeatedly use the unique Stokes-Helmholtz decomposition of any second-order tensor field, say $U$,  on a simply connected domain $\Omega$ given by the following statements:
\begin{equation}\label{SH_brief}
\begin{split}
U &= Gradz + \chi \ \ \mbox{on} \ \ \Omega\\
Div \chi &= 0 \ \ \mbox{on} \ \ \Omega\\
\chi . N & = 0 \ \ \mbox{on} \ \ \partial \Omega\\
Gradz . N & = U . N \ \ \mbox{on} \ \ \partial \Omega
\end{split}
\end{equation} 
The potentials are derived from a relation  analogous to the Helmholtz identity which shows there is  a  tensor field $A$  such that
\begin{equation}
\label{achi}
\chi=  Curl\, A ,\qquad Div\,A=0.
\end{equation}
In consequence, we have
\begin{equation}
\label{laA}
\Delta\,A = \ Curl\,\chi,
\end{equation}
where $\Delta$ is the Laplace operator.

The chosen boundary condition \eqref{SH_brief}$_3$ represents no loss and  is sufficient, for example, to recover classical linear elasticity theory in the absence of dislocations; see the end of Section \ref{mov}

\section{Plasticity implies  classical Volterra theory: an example}\label{plast-Volterra-Love}
In this Section a particular example is chosen  to illustrate the  connection between the classical Volterra and  plasticity formulations of dislocations.  The domains in which the classical Volterra problem is posed for a single dislocation and the corresponding one for plasticity theory  are different. In the set of points common to both domains, the stress in the Volterra problem is linearly related to the displacement gradient.
By contrast, stress in the plasticity  theory is linearly  related to
the difference between the total displacement gradient and the plastic distortion.
Consequently, it is important to establish what relationship, if any, exists  between the respective total displacement and stress fields.

Singularities occurring in the Volterra description considerably complicate the treatment of nonlinearities caused by evolving dislocation fields and  corresponding elastic distortion tensors. On the other hand,  \textit{a priori} singularities do not occur in the  plasticity theory for discrete dislocations. Their absence permits  realistic microscopic physics to be included
in the description of dislocation motion. We note that for the plasticity problem in the singular case, DeWit \cite{dewit1973theoryII, dewit1973theoryIII, dewit1973theoryIV} utilises results from the theory of distributions to derive explicit expressions for total displacement, elastic strain, and stress. Mura \cite{mura_micro} uses the eigenstrain distribution of a singular penny-shaped inclusion to form a body force in the usual way and then observes that the displacement solution based on the Green’s function approach gives exactly Volterra’s formula for the field of a dislocation; Eshelby \cite{eshelby1957determination} notes the correspondence as well for deriving the field of a dislocation loop. Kosevich \cite{kosevich1979crystal}  attempts a somewhat different explanation for the correspondence between the eigenstrain and Volterra formulations, which we have found to be ambiguous in its details. All of these explanations rely on explicit, Green's function-based formulae in homogeneous elastic media and none of them explain why the plasticity/eigenstrain formulation should recover the Volterra formulation as a limit; our analysis provides such an explanation. One example of the utility of our line of reasoning in this Section is presented at the beginning of Sec. \ref{scdisl} where a transparent,  qualitative explanation is provided for why it is natural to expect a difference in the result for uniqueness of stress fields of a specified dislocation density, in the traction-free case, between the quasi-static and dynamic cases (in a generally inhomogeneous elastic medium), without invoking any explicit formulae whatsoever.

The  example considered  concerns
a stationary  single straight line dislocation for which the  same Burgers vector is prescribed   for both the Volterra and plasticity dislocation theories.  Consequently, in this Section only,  dislocation densities  are  derived quantities and not data.   Moreover,  boundary value problems  in the Volterra theory may   involve   discontinuities and other singular behaviour.

\begin{figure}
\centering
\includegraphics[width=4.0in, height=3.5in]{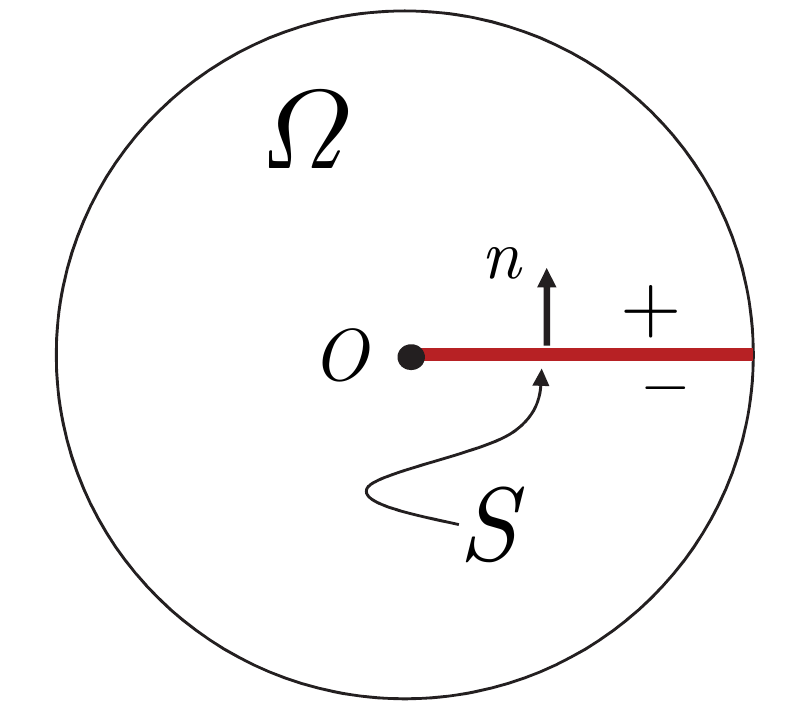}
\caption{Schematic of setting for Volterra dislocation.}
\label{fig:volterra_1}
\end{figure}

The  region  $\Omega$ in  Fig. \ref{fig:volterra_1}  denotes the unit disk in $\mathbb{R}^2$, whose centre  $O$ is the origin of a Cartesian rectangular coordinate system. The  region  $\Omega$ is  considered as the orthogonal  cross-section of a right circular cylinder with symmetry axis in the $x_3$ direction. Let $S$ denote the intersection of $\Omega$ with the  half-plane $x_{1}\geq 0$  within the $(x_1,x_3)$ plane; expressed otherwise, we have
\begin{equation}
\label{defS}
S = \{ (x_1, x_2)\in\Omega: x_1 \ge 0, x_2 = 0\}.
\end{equation}

Consider  the Volterra dislocation problem of  a static single straight dislocation along the $x_3-$axis under zero boundary tractions and no external body-force. On the slit region $\Omega\backslash S$ let the map $u: \Omega \backslash S \rightarrow  \mathbb{R}^3$ represent the   total  displacement field 
that possesses the following limits  as adjacent sides of $S$ are approached:
\begin{equation}\label{volterra0}
\begin{split}
& u^+(x_1) :=lim_{x_2\rightarrow 0^+} \, u(x_1, x_2),\qquad x_1 > 0, \\
& u^-(x_1) :=lim_{x_2\rightarrow 0^-} \, u(x_1, x_2),\qquad x_1 > 0, \\
& (Grad\, u)^+(x_1) := lim_{x_2\rightarrow 0^+} \, Grad\, u(x_1, x_2), \qquad x_1 > 0,\\
& (Grad\, u)^-(x_1) := lim_{x_2\rightarrow 0^-} \, Grad\, u(x_1, x_2), \qquad x_1 > 0.
\end{split}
\end{equation}

We seek to determine   the  map $u$ that satisfies
\begin{equation}
\label{volterra}
 Div\,  C (Grad\, u )^s = 0,\qquad x\in \Omega \backslash S,
\end{equation}
subject to the conditions
\begin{eqnarray}
\label{voltb}
 \left[ u \right]_{S}&:=& u^+ - u^-  = b, \qquad x_{1}>0,\\
\label{volttr}
 \left[ C (Grad\, u )^s \right]_{S}.n &:=& (C (Grad\, u^+)^s - C (Grad\, u^-)^s). n  = 0 \ \ \mbox{for} \ \  x\neq 0,\\
\label{voltbdry}
 C (Grad\, u)^s .N &=& 0 \ \ \mbox{on} \ \partial \Omega.
\end{eqnarray}
Here, $b \in \mathbb{R}^3$ is a given \emph{constant} Burgers vector, $\left[,\right]_{S}$ represents the jump across $S$,  $N$ is the unit outward vector normal   field on $\partial \Omega$ and $n=e_2$ is the unit vector normal on $S$.

Any such solution must satisfy  $|Grad\, u (x_1,x_2)| \rightarrow \infty$ as $(x_1,x_2) \rightarrow (0,0)$, since the line integral  of the displacement gradient taken anti-clockwise along any circular loop of arbitrarily small radius encircling the origin and starting from the ``positive'' side  of $S$ and ending at the ``negative''  side  must recover the finite value $-b$. This also implies that the displacement gradient field must diverge as $r^{-1}$ as $r \rightarrow 0$, where $r(x_1,x_2)$ is the distance of any point  $(x_{1},x_{2})$ from the origin.   Consequently,  the linear elastic energy density is not integrable for bounded bodies  that contain the origin.

\begin{figure}
\centering
\includegraphics[width=3.8in, height=3.5in]{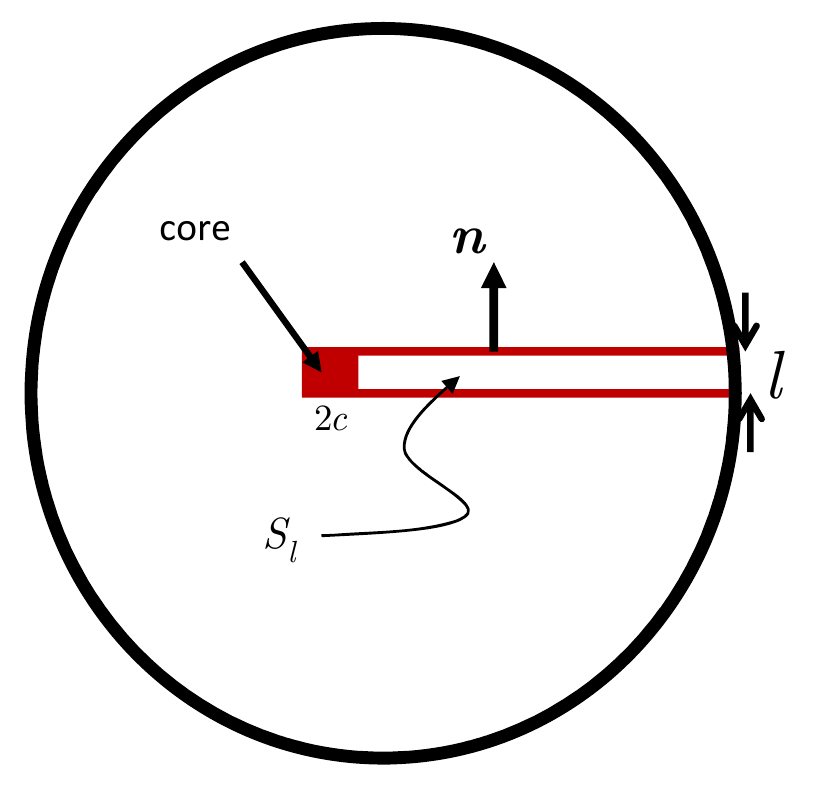}
\caption{Schematic of setting for dislocation in plasticity theory.}
\label{fig:volterra_2}
\end{figure}

With reference to Fig. \ref{fig:volterra_2}, we now introduce the ``slip region'' given by
\begin{equation}\label{Sl}
S_{l}=\left \{(x_1,x_2)\in \Omega \ | \ x_1\geq 0, \ |x_2| \leq \frac l2\right \}, \ l>0.
\end{equation}
and the ``plasticity core'',  a rectangular neighbourhood of  the dislocation defined as
\[
S^{l,c} = \left \{(x_1,x_2)\in \Omega \ | \ |x_1| < c, \ |x_2| \leq \frac l2\right \}.
\]

The plasticity core in the limit $l \rightarrow 0$ represents the line segment
\[
\left \{(x_1,x_2)\in \Omega \ | \ |x_1| < c, x_2 = 0 \right\}.
\]

The boundary value problem \eqref{volterra0}-\eqref{voltbdry} for the Volterra dislocation is defined on $\Omega\backslash S$.  In the plasticity theory of dislocations, however,  the boundary value problem is defined on the whole of $\Omega$ and 
is specified  by
\begin{equation}\label{plasticity}
\begin{split}
& Div \,C( U^{(E)})^s  = 0, \qquad x\in \Omega,\\
& C\, (U^{(E)})^s. N = 0, \qquad x\in \partial \Omega,
\end{split}
\end{equation}
where $U^{(E)}$,  the elastic distortion tensor field,  is smooth on $\Omega$. 

   The \textit{plastic distortion tensor field} $U^{(P)}$ is  defined as the difference between the gradient of the total displacement, denoted here to avoid confusion  by $u^d: \Omega \rightarrow \mathbb{R}^3$,  and the elastic distortion tensor $U^{(E)}$. 
Further physical motivation for  these field variables  will be presented in Sections \ref{stat} and \ref{qsbvp}.  Accordingly, on noting  that $u^d$ 
maps the whole of $\Omega$, we have
\begin{equation}
\label{prerel}
U^{(P)}(x)=Grad\,u^{d}(x)-U^{(E)}(x),\qquad x\in\Omega.
\end{equation}

The  explicit form selected for the plastic distortion tensor\footnote{For ease of presentation in this example, we adopt this discontinuous form for $U^{(P)}$. However, we note that  standard mollification of  $U^{(P)}$ can be used to produce a smooth approximating sequence of plastic distortion tensors to which the remaining arguments in this section may be applied.}, taken as data, is  given by
\begin{equation}
\label{Uph}
\begin{split}
& U^{(P)}(x) = \begin{cases}
&         g(x_1)\, \frac{1}{l}\, ( b \otimes n) \ \ \mbox{in} \ \  S_l,\\
&         0 \ \ \mbox{in} \ \ \Omega\backslash S_l,\\
         \end{cases}\\
\end{split}
\end{equation}
where $b=(b_{1},b_{2}, b_{3})$ is the constant Burgers vector,  $n \in \mathbb{R}^2$ is the unit vector normal to the `layer' $S_{l}$, $g(x_1) = 1$ for $x_1 \geq c > 0$, $g(x_1) = 0$ for $x_1 \leq -c$, and $g$ is a  monotone increasing function in $(-c,c)$.
 Hence, the non-uniformity of  $g(x_{1})$ is confined to the plasticity core region $S^{l,c}$. We remark that $n=e_{2},\, b=b_{1}e_{1}+b_{2}e_{2}$ for an edge dislocation, while $n=e_{2},\,b=b_{3}e_{3}$ for a screw dislocation. The parameters $l$ and $c$ are significant  in the physical modelling of dislocations: $l$ represents the interplanar spacing of a crystal and $c$ represents the non-vanishing core width of a crystal dislocation. Both $l$ and $c$ are observable quantities. From this point of view, the Volterra dislocation is an approximation (a large length-scale limit) of physical reality. 

The component in the $x_{1}-$direction  of each row of $U^{(p)}$  given by \eqref{Uph} is zero while their normal component along $n=e_2$ 
has a 
derivative in the $x_{1}-$direction that is non-zero only in the core region $S^{l,c}$. Therefore,   $Curl \, U^{(P)} =: -\alpha$ is non-vanishing only in the core. (Jumps in $U^{(P)}$ in the normal direction across the layer $S_{l}$ are not sensed by the (distributional) $Curl$). It follows from  Stokes' Theorem that
\begin{equation}
\label{Upj}
\int_Q Curl\, U^{(P)} \, e_3 \, dS = \int_{\partial Q} U^{(P)}. ds = b,
\end{equation}
for any area patch $Q$ that completely covers the plasticity core, $S^{l,c}$,  and whose  closed bounding curve $\partial Q$ intersects the layer $S_l$ in points with   $x_1$-coordinate greater than $c$. In the above, $e_3$ is the unit normal in the direction out of the plane in a right-handed sense.

Since the Volterra problem is posed on the  region  $\Omega \backslash S$, we seek to establish its equivalence with the plasticity problem on $\Omega\backslash S_l$ as $l\rightarrow 0$ and $c \rightarrow 0$.


The following  orthogonal Stokes-Helmholtz  decomposition holds for  the field $U^{(P)}$:
\begin{equation}
\label{Up_SH}
 U^{(P)} = Grad\, z - \chi,\qquad x\in\Omega,\\
\end{equation}
where the second order tensor field $\chi$ satisfies
\begin{equation}
\label{shchi}
\begin{split}
& Curl\, \chi = \alpha = - Curl\, U^{(P)},\qquad x\in\Omega,\\
& Div \,\chi = 0,\qquad x\in\Omega,\\
& \chi. N = 0 \qquad x\in \partial \Omega.
\end{split}
\end{equation}
This structure has the important implication that $\chi$ is a  continuous field on $\Omega$, \emph{for all} values of $l \ge 0$ and $c > 0$.\footnote{The mollification mentioned in Footnote 1 results in $\alpha$ and $\chi$ becoming smooth fields on $\Omega$.}
When $c = 0$,  $\chi$ becomes a  continuous field on the punctured domain $\Omega \backslash \{(0,0)\}$.

Let $\gamma(s)$ denote the line segment
\begin{equation}
\label{lseg}
\gamma (s) = (x_1,0) +  \left( -\frac{l}{2} + s l \right) n, \qquad s\in [0,1],  \ \mathrm{for \ fixed \ } \ x_1 > c.
\end{equation}
By virtue of     \eqref{Upj} and the  continuity of $\chi$, the  integral along $\gamma(s)$ of both sides of (\ref{Up_SH})  yields
\begin{equation}\label{z_jump}
lim _{l \rightarrow 0}  \int_{\gamma(s)}Grad\, z \,ds =\left[ z \right]_{x_1 > c}= z(x_1,0^+)_{x_1 > c} - z(x_1, 0^-)_{x_1 > c}  =lim _{l \rightarrow 0} \int_{\gamma(s)} U^{(P)}. ds = b  \ .
\end{equation}

\noindent
The system  (\ref{plasticity}) can be rewritten as
\begin{equation}
\label{rep410}
\begin{split}
&u^1 := u^d - z,\qquad x\in \Omega,\\
& U^{(E)} =Grad\,u^{1}+\chi,\qquad x\in \Omega,\\
& Div \, C\left( Grad\, u^1\right)^s  = - Div \,C\left( \chi\right)^s,\qquad x\in \Omega,\\
&   C\left( Grad\, u^1\right)^s .N  = -  C\ \left( \chi\right)^s .N, \qquad x\in \partial \Omega,
\end{split}
\end{equation}
and we note that  for $c > 0$,  $u^1$ is a  smooth field away from the core on $\Omega$ \emph{for all} values of $l  \ge 0$ (because $\chi$ is continuous and piecewise-smooth on $\Omega$).
 By integrating both sides of the expression $Grad\, u^d - Grad\, z = Grad\, u^1$ along the line segment  $\gamma(s)$, taking the limit $l \rightarrow 0$, and appealing to \eqref{z_jump}, we obtain
\[
\left[ u^d \right]_{x_1 > c}: =u^d (x_1, 0^+) - u^d(x_1, 0^-)  = \left[ z \right]_{x_1 > c} = b.
\]

The  continuity of $\chi$  and  the fact that $u^1$ is a solution to the system \eqref{rep410} for such a $\chi$ implies that for given $l\geq 0,\,c>0$, the total energy of the body is bounded; that is, $\frac{1}{2}\int_{\Omega} U^{(E)}:C\,U^{(E)}\,dx <\infty.$ 
Moreover,   these properties also imply that for $c>0$ the tractions in the plasticity formulation are always continuous on any internal surface of $\Omega$. In particular, we have
\begin{equation}
\nonumber
\left[ C \, \left( U^{(E)} \right)^{s}\right]_{x_{1}>c} n: = C\,\left( U^{(E)}\right)^{s}(x_1, 0^+) n - C\,\left( U^{(E)}\right)^{s}(x_{1}, 0^-) n = 0, \ \ \mbox{for} \ \ c > 0.
\end{equation}
which is valid for points even in the plasticity core $S^{l,c}$.

Upon noting that   $U^{(P)} = 0$ in $\Omega \backslash S_l$  and also that $S_l \rightarrow S$ as $l \rightarrow 0$,  we  recover the following relations  for $c > 0$,
\begin{equation}
\label{plsys}
\begin{split}
& Div \, C\, \left( Grad\, u^d \right)^s = 0, \qquad x\in\Omega \backslash S,\\
& \left[ u^d \right]_{x_{1} > c} = b,\\
& \left[  C\,\left( Grad\, u^d \right)^{s}\right]_{x_{1}>c} n:=\left\{C\, \left(Grad\, u^d(x_1, 0^+)\right)^s - C \left( Grad\, u^d(x_{1}, 0^-))^s\right)\right\}. n  = 0, \\
&  C ( Grad\, u^d)^s. N = 0, \qquad x\in\partial \Omega.
\end{split}
\end{equation}
The system \eqref{plsys} is formally identical to the  ``Volterra'' system \eqref{volterra0}-\eqref{voltbdry}.   It is in this sense that  the plasticity solution is equivalent to  the solution for the  classical Volterra dislocation problem. 

 The plasticity solution is a very good approximation to the Volterra solution in $\Omega \backslash S_l$ even for small $l > 0$ (compared to the radius of the body).   Comparison of finite element approximations \cite{zhang2015single} with the exact Volterra solution outside the plasticity core $S^{l,c}$ confirms that  within $S_l$ and elsewhere the correspondence is excellent.


We have thus explained  how a stationary solution to a Volterra dislocation problem may be regarded as the limiting form of solutions to a sequence of plasticity dislocation problems  having particular  plastic distortion tensors. We believe that the plasticity formulation is more general, practically versatile, and better  able to  deal with  dislocations in elastic solids, especially those that are  evolving. In the following sections we  prove certain  uniqueness results for the plasticity dislocation theory   that directly apply to  a body with an arbitrary collection of dislocation lines.

 The discussion of    the relationship between the Volterra and plasticity formulations has  assumed that   the plastic distortion is  data.  The treatment, however,  in the next three sections 
adopts the dislocation density tensor, and not the plastic distortion tensor, as data and shows that  the  plastic distortion tensor is not always  uniquely determined when this is the case.

\section{Stationary  (equilibrium)  problem}\label{stat}

The  simply connected  region  $\Omega$, which we recall is occupied by a self-stressed linear inhomogeneous anisotropic compressible elastic material  in equilibrium under zero applied body-force, prescribed dislocation density $\alpha$,  and non-zero surface traction,  is  adopted as the reference configuration. The  primary concern of this Section is to determine the self-stress occurring in $\Omega$, subject  to  relations \eqref{bdis}-\eqref{preeqeqn}, and to explore uniqueness issues. 
 The appropriate traction boundary value problem is stated as
\begin{eqnarray}
\label{ddelast}
\alpha&=&Curl\, U^{(E)},\qquad x\in\Omega,\\
\label{equi}
Div\,\sigma&=&  Div\,C\left(U^{(E)}\right)^{s} =0,\qquad x\in\Omega,\\
\label{tbvp}
\sigma.N&=& C\left(U^{(E)}\right)^{s}.N  = g,\qquad x\in\partial\Omega,
\end{eqnarray}
where $g(x)$ is a prescribed statically admissible surface traction vector (i.e., the resultant of the surface force and moment arising from the traction distribution is zero).

\subsection{Uniqueness of stress and elastic distortion}\label{statstedis}
We  prove for given  dislocation density $\alpha$ and  surface traction $g$ that the stress and the elastic distortion (up to a constant skew tensor) are unique.

\begin{proposition}\label{uniqueness_statics} 
In the traction boundary value problem (\ref{ddelast})-(\ref{tbvp}) for specified elastic modulus tensor $C(x)$, the stress tensor is uniquely determined by the prescribed dislocation density $\alpha(x), \,x\in\Omega$ and surface traction $g(x),\,x\in\partial\Omega$. The elastic distortion tensor $U^{(E)}$  is unique up to a constant skew-symmetric tensor field on $\Omega$.
\end{proposition}

\textbf{Proof}.  (Stated in \cite[Sec. 5]{willis1967second} for infinite regions subject to asymptotically vanishing stress at large spatial distance.)


Kirchhoff's uniqueness theorem \cite{kp71} 
is not immediately applicable and we proceed as follows. Let $U^{(E)}_{1},\,U^{(E)}_{2}$ be  
solutions to (\ref{ddelast})-(\ref{tbvp}) with $\widehat{U}^{(E)}:= U^{(E)}_1 - U^{(E)}_2$.  We have
\begin{eqnarray}
\nonumber
 \widehat \sigma& :=&\sigma_{1}-\sigma_{2}\\
\label{unss}
& =&C\left(\widehat U^{(E)}\right)^{s},\qquad x\in\Omega.
\end{eqnarray}

\noindent
But $U^{(E)}_{1},\,U^{(E)}_{2}$ each satisfy  relation (\ref{ddelast}) for prescribed dislocation density $\alpha$, and consequently
\begin{equation}
\label{diffelten}
Curl\, \widehat U^{(E)}=0,\qquad x\in\Omega,
\end{equation}
from which  we  infer the existence of a vector function $\widehat z$ such that $\widehat U^{(E)}=Grad\, \widehat z,\,x\in\Omega$,  since $\Omega$ is simply connected. It follows that  $\widehat \sigma=C \left(Grad\,\widehat z\right)^s$  satisfies a zero-traction boundary condition, and uniqueness is implied by Kirchhoff's theorem. Hence, the elastic distortion $U^{(E)}$ is unique to within a constant skew-symmetric tensor field on $\Omega.$

 The solution to the traction boundary value problem (\ref{ddelast})-(\ref{tbvp}) for spatially uniform elasticity and unbounded regions may be obtained   using either Green's function (see, for example, \cite{willis1967second,kgbb79}), or  Fourier transform techniques (see, for example, \cite{s51,s72, efs56}), or   stress functions as developed in \cite{k58}. Of course, the most practically efficient method for solving the system (\ref{ddelast})-(\ref{tbvp}) in  full generality uses  approximation techniques based on the finite element method described in  e.g.,  \cite{jiang2013least}    (cf. \cite{roy2005finite}). Related  convergence results and error estimates also are available.  Kr\"{o}ner's approach \cite{k58}    (with given dislocation density), even when applied to unbounded regions and homogeneous isotropic elasticity,  shows that the  stress and elastic \textit{strain}(i.e., symmetric part of $U^E$) are unique but that the skew symmetric part of the elastic distortion remains undetermined. Unlike  linear elasticity, the skew symmetric part of the elastic distortion in the present context of incompatible linear elasticity can
be spatially inhomogeneous even if the  symmetric part  vanishes. Circumstances in which this may occur represent important physical configurations \cite{roy2005finite,brenner2014numerical}   e.g., stress-free dislocation walls.

 Another  method of solution for (\ref{ddelast})-(\ref{tbvp}) follows \cite{roy2005finite}  and writes $U^{(E)}$ as a gradient of a vector field plus a tensor field that in general is not curl-free.    Both fields are then determined  from  equations (\ref{ddelast})-(\ref{tbvp}). (The decomposition is not exactly that of Stokes-Helmholtz and  is further discussed   at the end of this section).  The  component potential functions  of $U^{(E)}$ exist by explicit construction using  standard methods in  potential theory and elasticity theory.
We seek a solution of the form
\begin{equation}
\label{sheq}
U^{(E)}= - Grad\, z^{(E)}+ \chi^{(E)},\qquad x\in\Omega,
\end{equation}
where the function $\chi^{(E)}$  satisfies
\begin{eqnarray}
\label{ddch}
Curl\,\chi^{(E)}&=& \alpha,\qquad x\in\Omega,\\
\label{divch}
Div\,\chi^{(E)}&=& 0,\qquad x\in\Omega,\\
\label{bcch}
\chi^{(E)}.N &=& 0,\qquad x\in\partial\Omega.
\end{eqnarray}

The vector function $z^{(E)}$ in \eqref{sheq}  is then chosen to satisfy the system  obtained on elimination of $U^{(E)}$ between  (\ref{sheq}) and (\ref{equi}) and (\ref{tbvp}). We have
\begin{eqnarray}
\label{statz}
Div\,C\left(Grad\, z^{(E)}\right)^{s}  & =& Div\,C\left(\chi^{(E)}\right)^{s},\qquad x\in\Omega,\\
\label{statzshbc}
\left(C\left(Grad\,z^{(E)}\right)^{s}\right).N & =& - g+ \left(C\left(\chi^{(E)}\right)^{s}\right).N,\qquad x\in\partial\Omega.
\end{eqnarray}
Note that the tensor function $\chi^{(E)}$ is uniquely determined by  (\ref{ddch})-(\ref{bcch}) for prescribed $\alpha(x)$: for,
 let $\chi^{(E)}_{1},\,\chi^{(E)}_{2}$ be  
solutions. Define $\chi^{(E)}(x)=\chi^{(E)}_{1}(x)-\chi^{(E)}_{2}(x)$ so that  $Curl\, \chi^{(E)}(x)=0,\,x\in\Omega$. Therefore,  $\chi^{(E)}(x)=Grad\,\phi^{(E)}(x)$ for some twice continuously differentiable vector function $\phi^{(E)}(x)$ as $\Omega$ is simply connected.  On substitution in (\ref{divch}) and (\ref{bcch}), we conclude  that $\phi^{(E)}$ satisfies the harmonic Neumann boundary value problem
\begin{eqnarray}
\nonumber
Div\,Grad\,\phi^{(E)} & = & 0,\qquad x\in\Omega,\\
\nonumber
Grad\,\phi^{(E)}.N & = & 0,\qquad x\in\partial\Omega.
\end{eqnarray}
Consequently,  $\phi^{(E)}$ is constant  and therefore $\chi^{(E)}=0$.

Upon substitution of the uniquely determined  $\chi^{(E)}$  in (\ref{statz}) and \eqref{statzshbc}, we obtain a  linear elastic traction boundary value problem  for $z^{(E)}$ under non-zero body-force  and surface traction. It can  be verified  that the required necessary conditions are satisfied for the vanishing of the  sum of forces and moments due to the boundary load  $g$ and conditions on $\chi^{(E)}$. Uniqueness theorems in linear elastostatics state   that  $z^{(E)}(x)$ is unique to within an infinitesimal rigid body displacement. Hence, a solution $U^{(E)}$ to (\ref{ddelast})-(\ref{tbvp})   exists and is unique (up to a constant skew-symmetric tensor) by Proposition \ref{uniqueness_statics}.

Conversely, suppose that $U^{(E)}$ is a given solution of (\ref{ddelast})-(\ref{tbvp}). Then a 
 unique Stokes-Helmholtz decomposition of $U^{(E)}$ exists  given by
\begin{equation}
\label{Elshdec}
U^{(E)} = - Grad\, z + \chi,\qquad x\in\Omega,
\end{equation}
where $z,\,\chi$ respectively are sufficiently smooth vector and tensor functions that satisfy
\begin{eqnarray}
\label{statchi}
 Curl\,\chi&=& \alpha,\qquad x\in\Omega,\\
\label{statdiv}
Div\,\chi&=& 0,\qquad x\in\Omega,\\
\label{statchibc}
\chi.N &=& 0,\qquad x\in\partial\Omega,\\
\label{statzeq}
Div\,Grad\,z&=& - Div\,U^{(E)},\qquad x\in\Omega,\\
\label{statzbc}
Grad\, z.N &=& -U^{(E)}.N,\qquad x\in\partial\Omega.
\end{eqnarray}

Although the component potentials $\chi, \chi^{(E)}, z, z^{(E)}$ apparently 
satisfy different sets of governing equations, nevertheless, it is easily seen that the functions $z^{(E)}$ and $\chi^{(E)}$ satisfy the relations
\begin{equation*}
\begin{split}
& \chi = \chi^{(E)},  \ \ \  Grad\, z = Grad\, z^{(E)},\\
& z = z^{(E)} \ \ \mbox{up to a translation.}
\end{split}
\end{equation*}
  Notice   that as $U^{(E)}$ is uniquely determined in this stationary problem to within a skew-symmetric tensor, neither  the fields $Div\, U^{(E)}$ on the domain $\Omega$ nor $U^{(E)}.N$ on the boundary $\partial\Omega$ can be arbitrarily prescribed.

\begin{rem}[Reduction to  classical linear elasticity of the stationary problem]\label{statred}
We seek necessary and sufficient conditions for classical linear elasticity to be  recovered from \eqref{ddelast}-\eqref{tbvp}. A classical linear elastic solution in this context corresponds to \eqref{equi}-\eqref{tbvp} in which $Curl\, U^{(E)} = 0$.  It is straightforward to see that $\alpha = 0$ is the necessary and sufficient condition for $U^{(E)}$ satisfying \eqref{ddelast}-\eqref{tbvp} to be a classical linear elastic solution.
\end{rem}
\begin{rem}[Conditions in terms of the decomposition \eqref{Elshdec}]\label{stshred}

Necessary and sufficient conditions for  $U^{(E)}$ satisfying \eqref{ddelast}-\eqref{tbvp} to be a classical linear elastic solution may be expressed in terms of the potential functions occurring in \eqref{Elshdec}. When $Curl\, U^{(E)} = \alpha=0$ in \eqref{ddelast}-\eqref{tbvp}, \eqref{statchi}-\eqref{statchibc} necessarily give $\chi =0$. Also, the potential $z$ then satisfies $Div (C\,Gradz)^s = 0$ on $\Omega$ and $(C(Grad z)^s). N = -g$ on $\partial \Omega$.  Conversely, if $\chi = 0$ and the potential $z$ satisfies the conditions in the previous sentence, then $U^{(E)}$ defined by \eqref{Elshdec} is a classical linear elastic solution in this context, i.e.,  $Curl\, U^{(E)} = 0$ and \eqref{equi}-\eqref{tbvp} are satisfied.
\end{rem}

\subsection{Example: Stationary screw dislocation in the whole space}\label{stsc}
 The technique based on the Stokes-Helmholtz decomposition  applied to the stationary problem  in  Section~\ref{statstedis}  is illustrated by a simple example. Consider  the whole space occupied by a homogeneous isotropic compressible linear elastic material that contains a single stationary straight line screw dislocation located at the origin and directed along the positive $x_{3}$-axis. For simplicity,  no applied body-force acts, and  appropriate fields, including the Stokes-Helmholtz potential $z^{(E)}$,  asymptotically vanish to suitable order.  In particular, the traction $g(x)$ prescribed in \eqref{tbvp} vanishes in the limit as $x_{i}x_{i}\rightarrow\infty.$

The  dislocation density is specified to be
\begin{equation}
\label{scstdd}
\alpha(x)=|b|\delta(x_{1})\delta(x_{2})e_{3}\otimes e_{3},\qquad x\in \RR^{3},
\end{equation}
where we recall that  $\delta(.)$ represents the Dirac delta distribution,  and $e_{i},\, i=1,2,3,$ are  the unit coordinate vectors.  The multiplicative constant $|b|$ is selected to ensure that $|b|$ is the magnitude of the corresponding Burgers vector.  Without loss, all dependent field  variables  are assumed independent of  $x_{3}$ and to be of sufficient smoothness.

Consider the  decomposition \eqref{sheq}  for the elastic distortion tensor $U^{(E)}$. In view of relation \eqref{laA}, a tensor function $A^{(E)}$ exists that satisfies
\begin{equation}
\label{scexchi}
\chi^{(E)}=-Curl\,A^{(E)},\qquad Div\,A^{(E)}=0,\qquad x\in\RR^{2}.
\end{equation}
and
\begin{equation}
\nonumber
\Delta\, A^{(E)}(x)=\alpha(x),\qquad x\in\RR^{2}.
\end{equation}
Substitution from \eqref{scstdd} leads to
\begin{equation}
\label{Sca33}
\Delta\,A^{(E)}_{33}(x)=\alpha_{33}=|b|\delta(x_{1})\delta(x_{2}),\qquad x\in\RR^{2}.
\end{equation}

All other components of $A^{(E)}$ are harmonic in $\RR^{3}$ and are supposed to vanish asymptotically at large spatial distance. Therefore, they vanish identically by Liouville's Theorem.
The distributional solution to \eqref{Sca33} is given by
\begin{equation}\label{Aesoln}
A^{(E)}_{33}=\frac{|b|}{2\pi}\ln R,\qquad R^{2}=x_{\beta}x_{\beta},
\end{equation}
and in consequence  from \eqref{scexchi} the non-zero components of $\chi^{(E)}$ are
\begin{eqnarray}
\label{ch31fin}
\chi^{(E)}_{31}(x_{1},x_{2})&=& -\frac{|b|}{2\pi}\frac{x_{2}}{R^{2}},\\
\label{ch32fin}
\chi^{(e)}_{32}(x_{1},x_{2})&=& \frac{|b|}{2\pi}\frac{x_{1}}{R^{2}},
\end{eqnarray}
 which show that

\begin{eqnarray}
\nonumber
\left(Div\,\chi^{(E)}\right)_{\beta}&=& 0,\\
\nonumber
\left(Div\,\chi^{(E)}\right)_{3}&=& \chi_{31,1}+\chi_{32,2}\\
&=& \frac{|b|}{2\pi}\left[-\frac{2x_{1}x_{2}}{ R^{4}}+\frac{2x_{2}x_{1}}{R^{4}}\right]\\
\label{chzediv}
&=& 0,
\end{eqnarray}
and  (\ref{divch}) is satisfied in the sense of distributions. It can also be  verified  that the solution \eqref{Aesoln} satisfies $Div\, A^{(E)} = 0$, so that $\chi^{(E)}:= - Curl\, A^{(E)}$ implies $Curl \, \chi^{(E)} = \alpha$ from \eqref{Sca33}.

The vector function $z^{(E)}(x_{1},x_{2})$ satisfies  (\ref{statz}), the right side  of which by virtue of (\ref{divch}),  \eqref{ch31fin}, and \eqref{ch32fin}   becomes
\begin{eqnarray*}
\left[C\chi^{(E)}(x_{1},x_{2})\right]_{,j}&=& \left[\lambda \chi^{(E)}_{kk}\delta_{ij}+\mu\left(\chi^{(E)}_{ij}+\chi^{(E)}_{ji}\right)\right]_{,j}\\
&= &\mu\left(\chi^{(E)}_{ij,j}+\chi^{(E)}_{ji,j}\right)\\
&=& 0,
\end{eqnarray*}
where $\lambda$ and $\mu$ are the Lam\'{e} constants, and $\delta_{ij}$ is the Kronecker delta.

Consequently, $z^{(E)}(x_{1},x_{2})$ 
is the solution to the equilibrium equations of linear elasticity on the whole space. Assume that $z^{(E)}(x_{1},x_{2})$ is bounded as $R\rightarrow\infty$. Liouville's Theorem  implies that $z^{(E)}(x_{1},x_{2})$ is constant.  Accordingly,  by (\ref{sheq})  the non-zero  components of the asymmetric elastic distortion tensor  are
\begin{eqnarray}
\label{scel31}
U^{(E)}_{31}(x_{1},x_{2})&=& -\frac{|b|x_{2}}{2\pi R^{2}},\\
\label{scel32}
U^{(E)}_{32}(x_{1},x_{2})&=& \frac{|b|x_{1}}{2\pi R^{2}}.
\end{eqnarray}

Let $\partial\Sigma$ be the circle of radius $a$ whose centre is at the origin. The Burgers vector corresponding to the elastic distortion tensor whose non-zero components are \eqref{scel31} and \eqref{scel32} may be calculated from \eqref{beldis} and  gives $b=(0,0,b_{3})$ where
\begin{equation}
\nonumber
b_{3}=\oint_{\partial\Omega}U^{E)}_{3\beta}dx_{\beta}=|b|,
\end{equation}
 as previously stated.

Well-known  expressions (see,for example, \cite{hl82}) are easily derived  for the unique non-zero stress components, namely
\begin{eqnarray}
\label{strsta31}
\sigma_{31}(x_{1},x_{2})&=& -\frac{\mu |b|x_{2}}{2\pi R^{2}},\\
\label{strsta32}
\sigma_{32}(x_{1},x_{2})&=& \frac{\mu |b|x_{1}}{2\pi R^{2}}.
\end{eqnarray}

\section{The quasi-static boundary value problem}\label{qsbvp}
It is supposed that the dislocation density evolves as a prescribed tensor function of both space and time. The precise mode of evolution is unimportant for immediate purposes since the  dislocation density is adopted as  data.  The body is subject to specified applied time-dependent surface boundary conditions on  tractions and/or \emph{total displacements} (see (5.1)), although  the applied body-force   is assumed to  vanish (for simplicity and without loss of  generality).  The time-varying data  causes  the stress $\sigma(x,t)$ and elastic distortion $U^{(E)}(x,t)$  also  to be   time- dependent, and   the body  to change  shape with time.   Prediction of  the elastic distortion, stress, and change of shape   necessitates  introduction of the \textit{total  displacement field  $u(x,t)$}, \emph{that is identical to the field $u^d$ of Sec. 3, but which we henceforth refer to simply as $u$ for notational simplicity}. The corresponding  state-space consists of pairs $\left(u, U^{(E)}\right)$. We 
consider a re-parametrization of the state space, and for this purpose  recall that  \eqref{elplrelat}, namely
\begin{equation}
\label{eprel}
U^{(P)}:=Grad\, u - U^{(E)},\qquad (x,t)\in\Omega\times [0,T),
\end{equation}
is employed to define the plastic distortion tensor.  The set of pairs  $\left(u, U^{(P)}\right)$ then forms a new state-space.  Section \ref{plast-Volterra-Love} discusses  the 
connection between (\ref{eprel}) and the classical Volterra  mathematical model of dislocations.

Relation  (\ref{denel}) between the dislocation density and elastic distortion remains valid for  time evolution  problems, and in conjunction with (\ref{eprel}) leads to the formulae
\begin{eqnarray}
\nonumber
\alpha&=&Curl\, U^{(E)}\\
\nonumber
&=& Curl\,\left(Grad\,
 u-U^{(P)}\right)\\
\label{denpla}
&=&-Curl\,U^{(P)},\qquad (x,t)\in\Omega\times [0,T),
\end{eqnarray}
where $[0,T),\,T>0,$ is contained in the maximal interval of existence.

The constitutive relation (\ref{ssrel})  also remains valid and  expressed in terms of the plastic distortion becomes
\begin{eqnarray}
\nonumber
\sigma&=& C \left(U^{(E)}\right)^{s}\\
\label{qsss}
&=& C \left(Grad\, u -U^{(P)}\right)^s, \qquad (x,t)\in\Omega\times [0,T).
\end{eqnarray}

 The inertial term $\rho \ddot{u}$ is discarded  in the quasi-static approximation to the initial boundary value
problem.  The time-dependence, however,  of all other field variables is retained  with time serving  as a parameter.   The quasi-static boundary value problem, including (\ref{denpla}) repeated here for completeness,  at each $t\in [0,T)$, therefore  becomes
\begin{equation}
\label{denplarep}
\alpha =-Curl\,U^{(P)},\qquad (x,t)\in \Omega\times [0,T),
\end{equation}
and
\begin{equation}
\label{qsconste}
Div\,\sigma =0, \qquad (x,t) \in \Omega \times [0,T),
\end{equation}
or
\begin{equation}
\label{qseq}
  Div\,C\left(Grad\, u -U^{(P)}\right)^s = 0, \qquad(x,t) \in \Omega \times [0,T),
\end{equation}
subject to  either   traction boundary conditions
\begin{eqnarray}
\label{qseltbc}
\sigma.N
\label{qstbc}
&=& g,\qquad (x,t) \in \partial\Omega \times[0,T),
\end{eqnarray}
or  mixed boundary conditions
\begin{equation}
\label{qsmbc}
u= h,\qquad (x,t)\in\partial\Omega_{1}\times [0,T),
\end{equation}
\begin{equation}
\label{qsmtn}
\sigma.N  = g ,\qquad (x,t) \in\partial\Omega_{2} \times [0,T),
\end{equation}

where $\partial\Omega=\partial\Omega_{1}\cup\partial\Omega_{2}$, $\partial\Omega_{1}\cap\partial\Omega_{2}=\emptyset$. The vector functions $h(x,t)$ and $g(x,t)$ are prescribed.

The traction boundary value problem, specified by (\ref{denplarep}),  (\ref{qseq}) and (\ref{qstbc}), is formally identical to the stationary traction boundary value problem studied in Section~\ref{stat}.   We conclude that  the quasi-static stress tensor is uniquely determined while the  elastic distortion tensor is unique to within a skew-symmetric tensor.   Without modification, however, the previous argument cannot be applied to prove  uniqueness of either the plastic distortion tensor or the total displacement.

To investigate this aspect,    for each $t\in [0,T)$ let  the plastic distortion be completely represented by  its  Stokes-Helmholtz decomposition in the form
\begin{equation}
\label{qssh}
U^{(P)}= Grad\, z -\chi,\qquad (x,t) \in \Omega \times [0,T),
\end{equation}
where the incompatible smooth tensor potential $\chi$ satisfies the system
\begin{eqnarray}
\label{qsddch}
\alpha &=& Curl\, \chi,\qquad (x,t)\in\Omega \times [0,T),\\
\label{qsdiv}
Div\,\chi &=&0, \qquad (x,t)\in\Omega \times [0,T),\\
\label{qschbc}
\chi.N&=& 0,\qquad (x,t)\in\partial\Omega \times [0,T).
\end{eqnarray}

Similar comments contained in Section~\ref{statstedis}  regarding uniqueness of the tensor $\chi^{(E)}(x)$ and its vanishing with $\alpha(x)$  apply to the tensor $\chi(x,t)$ and the time-dependent dislocation density $\alpha(x,t)$. 

Define the vector functions $\widetilde{r}(x,t),\,\widetilde{s}(x,t)$  by
\begin{eqnarray}
\label{rtildedef}
\widetilde{r}&=& Div\,U^{(P)},\qquad (x,t)\in\Omega\times [0,T),\\
\label{stilde}
\widetilde{s}&=& U^{(P)}(x,t).N,\qquad (x,t)\in\partial\Omega\times [0,T).
\end{eqnarray}

In terms of  the continuously differentiable vector function $z(x,t)$ appearing in (\ref{qssh}) for each $t\in [0,T)$, these definitions are  equivalently expressed as
\begin{eqnarray}
\label{zeq}
Div\,Grad\,z &=& \widetilde{r},\qquad (x,t) \in \Omega \times [0,T),\\
\label{zbc}
Grad\,z. N&=& \widetilde{s},\qquad (x,t) \in \partial\Omega \times [0,T).
\end{eqnarray}
The vector -valued functions $\widetilde{r}(x,t),\,\widetilde{s}(x,t)$   at each time instant are restricted by the compatibility condition
\begin{equation}
\label{rscomp}
\int_{\Omega}\widetilde{r}\,dx=\int_{\partial\Omega}\widetilde{s}\,dS,
\end{equation}
but otherwise may be arbitrarily selected.  Here, we consider them as data,  along with the dislocation density tensor.

\begin{rem}
In the full stress-coupled theory of dislocation mechanics as a non-standard model within the structure of classical plasticity theory\cite{ma63,kosevich1979crystal,acharya2015dislocation,zhang2015single}, physically well-motivated and, in principle, experimentally observable evolution equations for the dislocation density and the plastic distortion arise naturally. There is of course some redundancy between the specification of both ingredients, and in the above models this is achieved in a self-consistent manner. On the other hand, as already demonstrated, the stationary traction boundary value problem of dislocation mechanics ((\ref{ddelast})- (\ref{tbvp})) is well-posed simply through the specification of the dislocation density. It is then reasonable to ask 
what extra \emph{minimal} ingredients beyond the specification of the dislocation density are
required to have a well-posed model of plasticity arising from the evolution of dislocations. This question is among  our primary concerns, without regard to the ease with which these minimal, extra ingredients can be physically determined.
\end{rem}

For specified $\widetilde{r}(x,t),\,\widetilde{s}(x,t)$,  the solution to the Neumann boundary value problem (\ref{zeq}) and (\ref{zbc}) is unique to within an arbitrary 
vector function of time $\widetilde{d}(t)$, and may be obtained   by any standard method in potential theory.




The results derived so far in this Section are summarised in the next Proposition.

\begin{proposition}\label{prop_5_1}
Consider the  plastic distortion tensors $U^{(P)(1)}(x,t),\,U^{(P)(2)}(x,t)$ that correspond to the same dislocation density $\alpha(x,t)$, and possess the same divergence and surface normal  components.   On appealing to the respective Stokes-Helmholtz decompositions, our previous results show that $U^{(P)(1)}=U^{(P)(2)}$. Consequently, specification of $\alpha,\,\widetilde{r},\,\widetilde{s}$ uniquely determines the plastic distortion.
\end{proposition}

We examine the implications of supposing that $\widetilde{r}, \widetilde{s}$  are arbitrarily assigned
but still subject to a prescribed dislocation density.
In the same manner as  previously shown, the dislocation density   uniquely determines  the tensor $\chi$ in the Stokes-Helmholtz decomposition (\ref{qssh}), but  arbitrary prescription of the vector functions $\widetilde{r}(x,t),\,\widetilde{s}(x,t)$ means that    $Grad\,z(x,t)$ remains  indeterminate.  Thus,
\begin{rem}\label{rem_5_2}
The quasi-static problem of  moving dislocation fields  with non-zero  dislocation density data admits an inevitable fundamental structural ambiguity pivotal  to the discussion of uniqueness.
\end{rem}
The  ambiguity is further explored in Section~\ref{mov}  devoted to  moving dislocations  subject to material  inertia.

We 
describe a slightly different proof to that in Proposition \ref{uniqueness_statics} to establish uniqueness of the
stress and elastic distortion in the quasi-static traction boundary value problem. Substitution of (\ref{qssh}) in (\ref{qseq}) and (\ref{qstbc}) yields
\begin{equation}\label{qsuzeq}
Div\,C \left(Grad\,(u-z) + \chi\right)^{s}  = 0, \qquad (x,t) \in \Omega \times [0,T),\\
\end{equation}
\begin{equation}
\label{qsuztbc}
C\left(Grad\,(u-z)\right)^{s}. N = g - \left(C\left(\chi\right)^{s}\right).N,\qquad x\in\partial\Omega\times [0,T).
\end{equation}

The tensor function $\chi$ appearing in these expressions is uniquely determined by the dislocation density $\alpha$. In consequence, the Kirchhoff uniqueness theorem of linear elastostatics ensures that $(u-z)$ is uniquely determined  by the system (\ref{qsuzeq}) and (\ref{qsuztbc}) to within an arbitrary rigid body displacement irrespective of the choice of $\widetilde{r},\,\widetilde{s}$.  This enables us to further conclude  that (\ref{eprel}), rewritten as
\begin{equation}
\nonumber
U^{(E)}=Grad\,\left(u-z\right)+\chi, \qquad (x,t) \in \Omega \times [0,T),
\end{equation}
implies the uniqueness of the elastic distortion (up to a constant skew-symmetric tensor field). Furthermore, the stress, given by
\[
\sigma = C \left( Grad\,\left(u-z\right) + \chi \right)^{s},\qquad (x,t) \in \Omega \times [0,T),
\]
is  unique.

  The vector functions  $\widetilde{r},\,\widetilde{s}$  uniquely determine $z$  to within an arbitrary  vector function of time only. Since it has just been shown that $(u-z)$  is unique to within an arbitrary rigid  body displacement,  the total displacement $u$ is also unique to within an arbitrary rigid body displacement dependent on  time as a parameter. Uniqueness is lost once $\widetilde{r},\,\widetilde{s}$ are arbitrarily chosen.

In the mixed boundary value problem, and also the displacement boundary value problem for which $\partial\Omega_{2}=\emptyset$, prescription of the boundary term $h$ requires that  $u$ and $z$ are separately considered.  The system (\ref{qsddch})-(\ref{qschbc})  still enables  $\alpha$ to uniquely determine $\chi$. However,    although 
$z(x,t)$ is uniquely determined to within an arbitrary vector $d(t)$ from (\ref{zeq}) and (\ref{zbc}), it still    inherits the  arbitrariness of $\widetilde{r},\,\widetilde{s}$.  Nevertheless, specification of  $\, \widetilde{r},\,\widetilde{s} \,$  leads to a unique  $Grad\,z$ which upon   insertion  into the system (\ref{qseq}), (\ref{qsmbc}), (\ref{qsmtn}) enables $u$ to be uniquely determined. Then  $U^{(E)}(x,t)$ can be calculated from (\ref{eprel}) and the stress from (\ref{qsss}).  However, like $z$, the field variables $U^{(P)}(x,t),\,u(x,t),\,U^{(E)}(x,t),\, \sigma(x,t)$  are ambiguous once  $\widetilde{r},\,\widetilde{s}$ become arbitrary.

It is of  interest to characterize the dependence of the fields  $U^{(E)}$ and $\sigma$ on $\widetilde{r},\,\widetilde{s}$ by rewriting (\ref{qseq}), (\ref{qsmbc}), (\ref{qsmtn}  as
\begin{eqnarray*}
 Div\, C\left[ (Grad\,(u -  z))^{s} \right] &=& - Div\, C\left( \chi \right)^{s}, \qquad (x,t) \in \Omega \times [0,T),\\
(u - z) &=& (h - z), \qquad  (x,t) \in \partial \Omega_1 \times [0,T),\\
 C\left[ (Grad\,(u -  z))^{s} \right].N  &=& g(x,t) - C\left( \chi \right)^{s}.N , \qquad (x,t) \in \partial \Omega_2 \times [0,T),
\end{eqnarray*}
where $z$ is obtained from \eqref{zeq} and \eqref{zbc} in terms of $\widetilde{r},\,\widetilde{s}$. It now follows that $(u - z)$, and consequently $Grad\, (u - z)$,  $U^{(E)},$ and the stress $\sigma$, depend on $\widetilde{r},\,\widetilde{s}$ only through the values of $z$ 
on  the boundary $\partial \Omega_1$. The arbitrariness of $z$ up to a vector function of time has no effect on the determination of either  $U^{(E)}$ or $\sigma$. In particular, the stress depends on $z$ through the term $Grad\,\left(u-z\right)^{s}$.

 These conclusions  are assembled  in the following Table, where the qualification to within appropriate rigid body displacements is understood.

\begin{center}
\begin{tabular}{|c||c||c||c|}\hline
\multicolumn{4}{|c|}{\bf  Specified dislocation density: Uniqueness in BVP's}\\
\hline
& {\it Traction BVP} & {\it Mixed BVP} & {\it All BVP:  $\widetilde{r},\,\widetilde{s}$ specified}\\ \hline\hline
$u$  & No & No & Yes\\ \hline
$U^{(P)}$ & No & No & Yes\\ \hline
$U^{(E)}$ & Yes & No & Yes\\ \hline
$\sigma$ & Yes & No & Yes  \\ \hline
\end{tabular}
\end{center}

\begin{rem}[Reduction to classical linear elasticity of quasi-static boundary value problem]\label{qsred}
We seek necessary and sufficient conditions for quasi-static classical linear elasticity to be recovered from \eqref{denplarep}-\eqref{qsmtn}. We define a pair of fields $(u, U^{(E)})$, or equivalently $(u, U^{(P)})$, satisfying \eqref{denplarep}-\eqref{qsmtn} as a classical linear elastic solution with stress given by $C\,(Grad \, u)^s$  provided $Curl\, U^{(E)}=0$ (or equivalently $Curl\, U^{(P)}=0$) and the total displacement $u$ satisfies the equations obtained from \eqref{qseq}-\eqref{qsmtn} on formally setting $U^{(P)}=0$. 

It is then easy to see that necessary and sufficient conditions for a solution $(u, U^{(P)}$) of \eqref{qseq},\eqref{qsmbc}, and\eqref{qsmtn} to be a classical linear elastic solution with stress given by $C\,(Grad \, u)^s$ are that
\begin{equation}\label{qs_reduct}
\begin{split}
Div\,C\left(U^{(P)}\right)^{s}&=0,\qquad x\in\Omega,\\
\left(C\left(U^{(P)}\right)^{s}\right).N&=0,\qquad x\in\partial\Omega_{2},\\
Curl \, U^{(P)} &= 0 \qquad x\in\Omega.
\end{split}
\end{equation}

The argument may be conducted in terms of  the Stokes-Helmholtz decomposition \eqref{qssh}. Since $Curl\, U^{(P)} = 0$ is equivalent to $\chi = 0$ on $\Omega$, it follows that   \eqref{qs_reduct} becomes
\begin{equation}\label{qs_reduct_SH}
\begin{split}
Div\,C\left(Grad\,z)^s\right)&=0,\qquad x\in\Omega,\\
\left(C \,(Grad\,z)^s\right).N&=0,\qquad x\in\partial\Omega_{2},\\
\chi &= 0 \qquad x\in\Omega.
\end{split}
\end{equation}
Consequently, \eqref{qs_reduct_SH} are the necessary and sufficient conditions for a classical linear elastic solution to be given by a triple $(u,\,\chi,\, z)$ that satisfies \eqref{denplarep}-\eqref{qsmtn}  and defines a pair $(u,\,U^{(P)})$ through  \eqref{qssh}.
\end{rem}


\section{Nonuniqueness for moving dislocations with material inertia}\label{mov}
We continue the discussion 
of a time evolving continuous dislocation distribution of specified density $\alpha(x,t)$, but now retain  inertia.
For  moving dislocations, the relation (\ref{denel}) and constitutive relations (\ref{ssrel}) together with (\ref{eprel})  continue to hold. The quasi-static equilibrium equation (\ref{qsconste}), however,  is replaced by the equation of motion \eqref{preeqmtn}, which  for convenience is repeated :
\begin{equation}
\label{mtn}
Div\,\sigma =\rho\ddot{u},\qquad (x,t)\in\Omega\times [0,T),
\end{equation}
where $\rho(x)>0$ is the positive mass density of the elastic body.

In terms of the total displacement vector  $u(x,t)$, for which initial Cauchy data is required, and plastic distortion tensor $U^{(P)}(x,t)$,  the initial boundary value problem studied in this Section is given by
\begin{equation}
\label{dyddpl}
\alpha= -Curl\, U^{(P)},\qquad (x,t) \in \Omega \times [0,T),
\end{equation}
\begin{equation}
\label{dynstr}
\sigma=C\left(Grad\,u-U^{(P)}\right)^{s},\qquad (x,t)\in\Omega \times [0,T),
\end{equation}
\begin{equation}
\label{dymtn}
 Div \,  C\left(Grad\, u - U^{(P)} \right)^{s} = \rho \ddot{u}, \qquad (x,t) \in \Omega \times [0,T),
\end{equation}
with displacement boundary conditions
\begin{equation}
\label{movdispbc}
u = h,\qquad (x,t)\in\partial\Omega_{1}\times (0,T),
\end{equation}

traction boundary conditions
\begin{equation}
\label{movtrbc}
 C\left(Grad\, u - U^{(P)} \right)^{s}.N =g, \qquad (x,t)\in\partial\Omega_{2}\times [0,T),
\end{equation}
and initial conditions
\begin{equation}
\label{movic}
u(x,0) = l(x),\qquad \dot{u}(x,0) =f(x),\qquad x\in \Omega,
\end{equation}
where $\partial\Omega =\partial\Omega_{1}\cup\partial\Omega_{2}$, , $\partial\Omega_{1}\cap\partial\Omega_{2}=\emptyset$, $\partial \Omega_2( t = 0) = \partial \Omega$ and $h(x,t),\,g(x,t),\,l(x),\,f(x)$ are prescribed functions.
\begin{rem}
When dealing with plasticity and dislocations, there is no natural reference configuration that can be chosen. The as-received configuration of the body may well be plastically deformed with respect to some prior reference and support a non-vanishing stress field. Moreover, it is not physically reasonable to require that some prior distinguished reference be known to determine the future evolution of the body and its state from the as-received one. Thus, the initial condition on the displacement is most naturally specified as $l = 0, x \in \Omega$ in \eqref{movic}. However, for many problems a displacement from a prior reference may be unambiguously known at $t =0$, e.g., when interrogating a motion of some reference from an intermediate state after some time has elapsed and considering the motion from this `intermediate' state now as the reference. It is for such situations that we allow for a general non-vanishing initial condition on displacement as in \eqref{movic}. 

Of note is also the fact that the formulation involves no displacement boundary condition at $t = 0$, allowing only the specification of tractions on the entire boundary at the initial time. These issues are further dealt with in Step 2 of the proof of Theorem \ref{uniq}.
\end{rem}
Unique specification of $U^{(P)}(x,t)$ implies that $u(x,t)$ is the solution to an initial mixed boundary value problem in linear elasticity subject to known body-force in (\ref{dymtn}), known boundary conditions (\ref{movdispbc}) and (\ref{movtrbc}), and known initial conditions (\ref{movic}).   Uniqueness of $u$ then follows from appropriate theorems in linear elastodynamics (see \cite{kp71}) and implies the unique determination of the elastic distortion $U^{(E)}(x,t)$ and stress $\sigma(x,t)$. However, the plastic distortion   $U^{(P)}(x,t)$ is not uniquely determined  from (\ref{dyddpl}) for given dislocation density $\alpha(x,t)$. Indeed, we have the following Theorem.

\begin{thm}[Non-uniqueness]\label{uniq}
 The   stress tensor $\sigma,$ elastic distortion tensor $U^{(E)}$, total displacement vector $ u,$ and  plastic distortion tensor $U^{(P)}$, belonging to the linear system  (\ref{dyddpl})-(\ref{movic})  are not in general  uniquely determined by the  prescribed data  $\alpha,\,h,\,g,\,l,\, \mbox{and }f$. Exceptions are noted below.
\end{thm}

\textbf{Proof}

The proof proceeds in three main steps and involves the Stokes-Helmholtz decomposition.

\textit{Step 1. Stokes-Helmholtz decomposition.}

Consider the relation (\ref{dyddpl}). The Stokes-Helmholtz decomposition    completely  represents $U^{(P)}$ as
\begin{equation}
\label{dysh}
U^{(P)}(x,t)= Grad\, z(x,t)-\chi (x,t),\qquad (x,t)\in\Omega\times [0,T),
\end{equation}
where the incompatible smooth tensor  potential function $\chi(x,t)$ satisfies the system  (\ref{qsddch})-(\ref{qschbc}). The  dislocation density $\alpha(x,t)$ therefore uniquely determines $\chi(x,t)$  for each $t\in [0,T)$. We recall that the boundary condition \eqref{qschbc} entails no loss and in particular ensures that $\chi$ vanishes with $\alpha.$

The boundary value problem (\ref{zeq})-(\ref{rscomp}) for the vector  potential  function $z(x,t)$ is replaced by the analogous system for the time derivative $\dot{z}(x,t)$ which accordingly for each $t\in [0,T)$ satisfies
\begin{eqnarray}
\label{zdoteq}
Div\,Grad\,\dot{z}&=& r,\qquad (x,t)\in \Omega \times [0,T),\\
\label{zdotbc}
Grad\,\dot{z}. N&=& s,\qquad (x,t) \in\partial\Omega \times [0,T),
\end{eqnarray}
for vector functions  $r(x.t),\,s(x,t)$  constrained at each time instant  to satisfy the compatibility condition
\begin{equation}
\label{movrscomp}
\int_{\Omega}r\,dx=\int_{\partial\Omega}s\,dS,\qquad t\in [0,T).
\end{equation}
As remarked in Section~\ref{qsbvp}, prescription of only the dislocation density $\alpha(x,t)$, but with the vector functions $r(x,t),\,s(x,t)$  arbitrarily ascribed,  creates structural ambiguities which are considered in Steps $2$ and $3$. Our aim is to identify  the essential  role of the fields $r,s$ in the determination of uniqueness.


When $r(x,t)$ and $s(x,t)$ are prescribed, the solution $\dot{z}(x,t)$ to the Neumann system (\ref{zdoteq})-(\ref{movrscomp}) is unique to within an arbitrary function of time, $d(t)$ for $t\in[0,T)$. Let $z^{(0)}(x)$ denote the initial value of $z(x,t)$. A time integration of the solution $\dot{z}(x,t)$ then shows that  $z(x,t)-z^{(0)}(x)$  is unique to within an arbitrary vector function of time, say $d_{1}(t)$. However, the  unique determination of stress and total displacement in the problem (\ref{dyddpl})-(\ref{movic}) depends  on the uniqueness of $U^{(P)}$,  which in turn  depends upon that of  $Grad\, z(x,t)$. Thus, the arbitrary vector function $d_{1}(t)$ is immaterial and can be ignored.



The next step is  to   calculate the initial terms $Grad\, z^{(0)}(x)$ and  $U^{(P)}(x,0)$.

\textit{Step 2. Initial physical data for $Grad\, z$ and $U^{(P)}$.}
It is a natural  requirement in  dislocation and plasticity theories that assigned initial data  should be observable in the current configuration and without the requirement of the knowledge a distinguished prior reference.

The evolution of $U^{(P)}$ depends on  its initial value in the as-received configuration of the body.  Thus, it is important to ascertain whether or not  initial values of  $U^{(P)}$ can be derived  from measurements conducted  on the body in the initial configuration. In this respect, we suppose that the dislocation density $\alpha(x,t)$ is a physically measurable observable for all $(x,t)\in \Omega\times [0,T)$ whose initial value $\alpha^{(0)}(x)$  is therefore  assumed  known.

We additionally suppose  that initial displacement, velocities, and accelerations can be physically measured so that we have
\begin{eqnarray}
\label{alphainit}
\lim_{t\rightarrow 0}{\alpha(x,t)} &=& \alpha^{(0)}(x), \qquad x\in\Omega,\\
\label{kininit}
u(x,0)&=& l(x),\quad \dot{u}(x,0)=f(x),\quad \ddot{u}(x,0)= m(x), \qquad x\in\Omega,
\end{eqnarray}
where $\alpha^{(0)}(x), \,l(x),\,f(x),\,m(x) $ are  known  from practical observation. \emph{As already mentioned,
$ l =0, x \in \Omega$ is expected to be a specification that is commonplace.} 

The initial value $U^{(P)(0)}(x)$ of $U^{(P)}(x,t)$ is  completely represented by the corresponding  Stokes-Helmholtz decomposition given by
\begin{equation}
\label{insh}
U^{(P)(0)}=Grad\, z^{(0)}-\chi^{(0)},\qquad x\in\Omega.
\end{equation}
By continuity, the equations of motion (\ref{dymtn}) and boundary conditions (\ref{movdispbc}) and (\ref{movtrbc}) are taken to   hold in the limit as $t\rightarrow 0^{+}$.

 In accordance with the previous treatment, the initial value $\chi^{(0)}(x)$ is  uniquely determined  from the system
\begin{eqnarray}
\label{inichdd}
Curl\,\chi^{(0)}&=&\alpha^{(0)},\qquad x\in\Omega,\\
\label{divinichi}
Div\,\chi^{(0)}&=&0,\qquad x\in\Omega,\\
\label{movchibc}
\chi^{(0)}.N&=& 0,\qquad x\in\partial\Omega,
\end{eqnarray}

Substitution  of (\ref{movic}) and \eqref{alphainit}-(\ref{insh})  in the equations of motion (\ref{dymtn}), assumed valid at $t=0$, leads to the equation for the initial value $z^{(0)}(x)$ of $z(x,t)$. We obtain
\begin{equation}
\label{zoeqn}
Div\, C \left(Grad\,\left(l-z^{(0)}\right) \right)^s  + Div\,  C\left(\chi^{(0)}\right)^s= \rho m,\qquad x\in\Omega,
\end{equation}
which after rearrangement becomes
\begin{equation}
\label{rezo}
Div\,C \left(Grad\,z^{(0)} \right)^{s} -\rho F=0, \qquad x\in\Omega,
\end{equation}
where the pseudo-body force $F(x)$ is uniquely defined to be
\begin{equation}
\label{bfo}
\rho F= Div\, C \left(Grad\, l  + \chi^{(0)} \right)^{s}  - \rho m, \qquad x\in \Omega.
\end{equation}

Moreover,   the traction  boundary condition may be written as
\begin{equation}
\label{movzic}
  C \left(Grad\, z^{(0)}\right)^{S}.N = -g(x,0) + \left[ C\left(Grad\, l+\chi^{(0)}\right)^{s} \,\right].N,\qquad x\in\partial\Omega_{2}.
\end{equation}

When  traction is specified everywhere on the surface so that $\partial\Omega_{2}=\partial\Omega$, then  $Grad\, z^{(0)}(x)$ is uniquely determined from (\ref{rezo}) and (\ref{movzic}) to within a constant skew-symmetric tensor field. The initial vector $z^{(0)}(x)$ is thus determined to within a rigid body displacement.


The specification of the function $g(\cdot, 0)$ on $\partial \Omega$ can, at times, involve the following considerations: Suppose that from some time $-t_{1} <0$ prior to time $t=0$ the body deforms subject to prescribed mixed boundary data. Then  the `reaction' tractions can be measured on that part $\partial\Omega_{1}$ of the boundary on which displacements are specified, and consequently 
 are known everywhere on $\partial\Omega$ at time $t=0$. In this sense, the mixed boundary value problem  can be replaced by a traction boundary value problem which as just shown uniquely determines  $Grad\, z^{(0)}$ to within a constant skew-symmetric tensor field.

 With the fields  $\chi^{(0)}$ and  $Grad\, z^{(0)}$ known, we obtain the initial plastic distortion $U^{(P)(0)}$ from \eqref{insh}. The time-evolution of $U^{(P)}$
represented by the decomposition \eqref{dysh} depends upon the calculation of
the time evolution of $\chi$ subject to  specified data $\alpha$, augmented by  the evolution of $z$ through integration of the solution to  (\ref{zdoteq})-(\ref{zdotbc}) for chosen $r$ and $s$. In consequence, for each definite choice of $r$ and $s$, $U^{(P)}$ is uniquely determined to within the arbitrary constant skew-symmetric  tensor present in the initial conditions for $U^{(P)}$.
Although this arbitrariness in $U^{(P)}$ does not affect the unique determination of the total displacement and stress from the system (\ref{dynstr})-(\ref{movic}), nevertheless,  the total displacement and stress are each  ultimately affected by the particular choice of $r$ and $s$.


\textit{Step 3. Conclusion of proof.}

Steps 1 and 2 establish  to within appropriate arbitrary constants, that the functions $\chi(x,t), \,\chi^{(0)}(x),$ and $Grad\, z^{(0)}(x)$ are uniquely determined  by the data  $\alpha(x,t),\,\alpha^{(0)}(x),\,$
\linebreak
$ l(x),\, m(x), \, h(x,t),\,h(x,0),\,g(x,t),$ and $ g(x,0)$.   As mentioned, however, the field $Grad\, z(x,t)$ is ambiguous due to the arbitrariness   of the vector functions $r(x,t)$ and $s(x,t)$.

The arbitrariness of $r,s$ also affects  the determination of  $U^{(P)}(x,t)$  from \eqref{dysh}, so that   terms dependent upon $U^{(P)}$  appearing  in \eqref{dymtn}  and the surface traction \eqref{movtrbc} create  indeterminacy  in the elastodynamic system   \eqref{dymtn}-\eqref{movic} for $u(x,t)$. In general,  the dislocation density does not uniquely determine  the total displacement $u(x,t)$,  and therefore also 
 the stress $\sigma(x,t)$ from \eqref{dynstr}. In order to identify  exceptions,  Lemmas~ \ref{unstress} and ~\ref{undisp} derive necessary and sufficient conditions for  the stress and  total displacement to be unique. Contravention  of these conditions provides sufficient and necessary conditions for  non-uniqueness of the respective variables.

Let $(r^{(\gamma)},\,s^{(\gamma)}),\,\gamma=1,2 $ be 
choices of $(r,s)$ that correspond to  the same prescribed dislocation density, boundary and initial conditions, and elastic moduli in (\ref{mtn})-(\ref{movic}).  Let
\begin{equation}
\label{diffpldi}
U^{(P)(\gamma)}= Grad\, z^{(\gamma)} -\chi^{(\gamma)}, \qquad (x,t) \in \Omega\times [0,T),
\end{equation}
be the Stokes-Helmholtz decomposition of the respective  plastic distortion tensors, where $z^{(\gamma)}(x,t), \, \gamma =1,2$   are each determined uniquely up to an additive vector function of time by the pair $(r^{\gamma},\,s^{\gamma})$. Corresponding initial values $z^{(0)(\gamma)}$, as just shown, are unique to within a rigid body displacement. Each $\chi^{(\gamma)}(x,t)$ is uniquely determined by the prescribed dislocation density and is unaffected by the choice of $(r,\,s)$. Consequently, $\chi^{(1)}=\chi^{(2)},\,(x,t)\in\Omega\times [0,T).$  By contrast,  $Grad\, z^{(\gamma)}$ is unaffected by the dislocation density, but is  uniquely determined   by the  given $(r^{(\gamma)},s^{(\gamma)})$. Set
\begin{eqnarray}
\nonumber
U^{(P)}& =&U^{(P)(1)}-U^{(P)(2)},\qquad (x,t) \in \Omega\times [0,T),\\
\label{graddiffz}
 Grad\, z&=& Grad\, z^{(1)}- Grad\, z^{(2)},\qquad (x,t) \in \Omega\times [0,T),
\end{eqnarray}
to obtain
\begin{equation}
\label{diffpldis}
U^{(P)}=Grad\,z,\qquad (x,t)\in\Omega\times [0,T),
\end{equation}
from which  follows
\begin{eqnarray}
\label{diffeldis}
U^{(E)}&=& U^{(E)(1)}-U^{(E)(2)}\\
\nonumber
&=& Grad\, u-U^{(P)}\\
\label{diffeuz}
&=& Grad\,(u-z),\qquad (x,t)\in\Omega\times [0,T),
\end{eqnarray}
where $U^{(E)(\gamma)} $ denotes the respective elastic distortions, and
\begin{equation}
\label{diffdispl}
  u(x,t)=u^{(1)}(x,t)-u^{(2)}(x,t)
\end{equation}
is the difference between the corresponding total displacements. We also denote the difference  stress by:
\begin{eqnarray}
\nonumber
\sigma&=&\sigma^{(1)}-\sigma^{(2)}\\
\nonumber
&=& C \left(U^{(E)(1)}\right)^{s}-C\left(U^{(E)(2)}\right)^{s}\\
\label{diffstre}
&=&C\left(U^{(E)}\right)^{s} = C \left(Grad\, (u - z)\right)^{s}, \qquad (x,t) \in \Omega\times [0,T).
\end{eqnarray}

On substituting for $U^{(P)(\gamma)}$ in the initial boundary problems (\ref{dymtn})-(\ref{movic}), and on taking  differences,  we obtain the system
\begin{eqnarray}
\label{movdiffmtn}
Div\, C\left(Grad\,(u-  z)\right)^s&=&\rho\ddot{u}, \quad (x,t)\in\Omega\times [0,T),\\
\label{diffmovbcs}
u &=&0,\quad (x,t)\in\partial\Omega_{1}\times [0,T),\\
\label{diffmovtbc}
  C\left(Grad\,(u-  z)\right)^s.N&=& 0,\quad (x,t)\in\partial\Omega_{2}\times [0,T),
\end{eqnarray}
\begin{equation}
\label{diffmovic}
u(x,0)=0,\qquad \dot{u}(x,0)=0,\qquad x\in\Omega.
\end{equation}



\begin{lem}\label{unstress}
The stress tensors  for the problems defined by (\ref{dyddpl}) -(\ref{movic}) corresponding to common data $\alpha,h,\,g,\,l,$ and $f$ but   different  plastic distortions $U^{(P)(1)}$ and $U^{(P)(2)}$ are identical if and only if 
$Grad\, z(x,t)$ defined by \eqref{graddiffz} for $(x,t) \in \Omega \times [0,T)$ and arising from the two plastic distortion fields, is at most a time-dependent spatially uniform skew-symmetric tensor field.

When $U^{(P)(1)}$ and $U^{(P)(2)}$ are  generated from  specified pairs $(r^{(\gamma)},\,s^{(\gamma)}),\,\gamma=1,2 $ (and  common dislocation density $\alpha$),  equivalent necessary and sufficient conditions for identical stress in the two problems are that $r^{(1)}=r^{(2)}$ and that $s^{(1)}$ and $s^{(2)}$  differ by at most   the cross-product of an arbitrary spatially independent vector field with the surface normal $N$  on the boundary $\partial \Omega$.

\end{lem}

\textbf{Proof}. \emph{Necessity}.  We assume that in (\ref{diffstre}), the difference stress identically vanishes so that  $\sigma(x,t)\equiv 0,\,(x,t)\in\Omega\times [0,T),$  and 
\begin{equation}
\label{diffzeruz}
C\left(Grad\,(u-z)\right)^s=0, \qquad (x,t) \in\Omega \times [0,T),
\end{equation}
which from \eqref{korn_proxy} implies
\begin{equation}
\label{zersuv}
u(x,t) =z(x,t) + a(t) +x \times \omega(t),\qquad (x,t)\in\Omega\times [0,T),
\end{equation}
where $a(t)$ and $\omega(t)$ are arbitrary vector functions of time $t$ alone. 
On the other hand, substitution of (\ref{diffzeruz}) in  (\ref{movdiffmtn}) shows that $\ddot{u}=0$ and therefore on using the initial conditions  (\ref{diffmovic}),
we  conclude that
\[
u(x,t) = 0, \qquad (x,t) \in\Omega\times  [0,T).
\]

Thus, we have
\begin{equation}\label{nec_suf_stress}
z_{i,j} (x,t) = - e_{ijk} \, \omega_{k}(t), \qquad (x,t) \in \Omega \times[0,T),
\end{equation}
i.e., $Grad\, z$ is a  time-dependent, spatially uniform skew-symmetric tensor field.

\emph{Sufficiency.} Assume $Grad\, z$ is a  time-dependent, spatially uniform skew-symmetric tensor field given by (\ref{nec_suf_stress}) and that (\ref{movdiffmtn})-(\ref{diffmovic}) are satisfied. Then, $\left(Grad\,z\right)^{s}=0$, and by Neumann's uniqueness theorem for linear elastodynamics (cp., \cite{kp71}), we conclude that
\[
u = 0, \qquad (x,t) \in \Omega\times [0,T).
\]
Substitution in (\ref{diffstre}) then implies that
\[
\sigma = 0, \qquad (x,t) \in \Omega\times [0,T),
\]
 and sufficiency is established.

The necessary and sufficient condition (\ref{nec_suf_stress}) for vanishing difference stress, where  $\omega(t)$ is an arbitrary function of time, may   alternatively be  expressed in terms of the difference functions $r=r^{(1)}-r^{(2)},\,s=s^{(1)}-s^{(2)}$.   Again, we first establish the corresponding necessary conditions which easily follow by  substitution of (\ref{nec_suf_stress}) in expressions corresponding to (\ref{zdoteq}) and (\ref{zdotbc}).  We obtain
\begin{eqnarray}
\label{diffr}
r(x,t)&=& 0,\qquad (x,t)\in \Omega\times [0,T),\\
\label{diffs}
s(x,t)&=& -N\times \omega^\circ (t),\qquad (x,t)\in\partial\Omega\times [0,T),
\end{eqnarray}
where $\omega^\circ$ is an arbitrary vector function of time. Note that the values of $r$ and $s$ given by  (\ref{diffr}) and (\ref{diffs}) satisfy the compatibility relation (\ref{movrscomp}).

For sufficiency,  assume that (\ref{diffr}) and (\ref{diffs}) hold  for an arbitrarily specified vector function of time $\omega^\circ$. Then, Neumann's uniqueness theorem for  the potential problem (\ref{zdoteq}) and (\ref{zdotbc}) subject to $r,\,s$ given by (\ref{diffr}) and (\ref{diffs}) yields
\begin{equation}\label{xxx1}
\dot{z}_{i,j} = -e_{ijk} \,\omega^\circ_k(t).
\end{equation}
Our hypothesis on the initial data and the considerations of \emph{Step 2}  show that the initial condition on the difference $Grad\, z$ can  be at most a spatially uniform skew-symmetric tensor field. This combined  with (\ref{xxx1}) implies that  to within an additive constant, $Grad\, z$ is of the form given by (\ref{nec_suf_stress}) for some vector function $\omega$ of time. Then \eqref{diffstre}-\eqref{diffmovic} imply $\sigma = 0$.

The proof of Lemma \ref{unstress} is complete.

\begin{lem}\label{undisp}
The total displacements $u^{(1)}(x,t)$ and $u^{(2)}(x,t)$ for the respective problems defined by (\ref{dyddpl}) -(\ref{movic}) subject to common data $\alpha,h,\,g,\,l,$ and $f$ and  plastic distortions $U^{(P)(1)}$ and $U^{(P)(2)}$ are identical if and only if the difference $Grad\, z(x,t)$ for $(x,t) \in \Omega \times [0,T)$, defined in (\ref{graddiffz}), satisfies
\begin{eqnarray}
\label{6.42}
Div\, C\left(Grad z\right)^{s}&=&0,\qquad x\in\Omega\times [0,T),\\
\label{6.43}
C\left(Grad\, z\right)^s.N &=& 0,\qquad x\in\partial\Omega_{2}\times [0,T).
\end{eqnarray}
\end{lem}

\textbf{Proof}. \emph{Necessity}. Let $u=u^{(1)}-u^{(2)}=0$. Then (\ref{movdiffmtn})-(\ref{diffmovic}) imply (\ref{6.42}) and (\ref{6.43}) for the difference vector $z(x,t)$.
Consequently, a necessary condition for  uniqueness of the total displacement  is that $U^{(P)(\gamma)}$, or alternatively $\left(r^{(\gamma)},\,s^{(\gamma)}\right)$, $\gamma = 1,2$, produce a solution to (\ref{zdoteq}) and (\ref{zdotbc})  compatible with a solution to   (\ref{6.42}) and (\ref{6.43}). Such solutions include a large class of vector fields $z$ whose gradient, $Grad\,z$, is non-trivial.


\emph{Sufficiency.} Suppose $z(x,t)$ satisfies  (\ref{6.42}) and (\ref{6.43}). Then (\ref{movdiffmtn}) -(\ref{diffmovic}) become
\begin{eqnarray}
\label{redmtn}
Div\, C \left(Grad\, u\right)^s &=&\rho\ddot{u},\qquad (x,t)\in\Omega\times [0,T),\\
\label{reddbc}
u(x,t)&=& 0,\qquad (x,t)\in \partial\Omega_{1}\times [0,T),\\
\label{redtbc}
C\left(Grad\, u\right)^s.N &=& 0,\qquad (x,t)\in\partial\Omega_{2}\times [0,T),\\
\label{redic}
u(x,0)&=&\dot{u}(x,0)=0,\qquad x\in\Omega.
\end{eqnarray}

The Neumann uniqueness theorem in linear elastodynamics states that at most only the trivial solution $u(x,t)\equiv 0$ exists to the system (\ref{redmtn})-(\ref{redic}), and consequently we have $u^{(1)}=u^{(2)}$.

The proof of Lemma \ref{undisp} is complete, and Theorem \ref{uniq} is established.

\begin{rem}\label{rem_6_1} The proof of Lemma \ref{unstress} shows that a necessary condition for the  stress to be unique  in   problems satisfying the hypothesis of Lemma \ref{unstress} is that the corresponding total displacements are identical. However, in the displacement and mixed problems, it is easily shown that violation at some time instant of the condition $s^{(1)}-s^{(2)}= (Grad\,\dot{z}).N= N\times a$ on $\partial \Omega_1$, where $a$ is a constant vector, is consistent with identical total displacements but not identical  stress. For the traction problem,  i.e., $\partial\Omega_{1} = \emptyset$, any solution  to (\ref{6.42})-(\ref{6.43}) necessarily satisfies (\ref{nec_suf_stress}),  and therefore the stress must be unique.
\end{rem}

Procedures described in this Section are illustrated by  the single screw dislocation uniformly moving in the whole space. Other treatments of the same problem include those presented in
 \cite{e53, ma63, lazar2009gauge}.

 Nevertheless,before proceeding,  we complete the discussion of  conditions for the reduction of  dislocation problems to  corresponding ones in classical linear elasticity. The development  employs the potentials appearing in the Stokes-Helmholtz decomposition \eqref{dysh}, and may be regarded as a
 special case ($\alpha=0$) of  Theorem~\ref{uniq}.

\begin{rem}[Reduction to classical linear elasticity for the dynamic initial-boundary value problem]\label{dynred1}
We seek necessary and sufficient conditions for dynamic classical linear elasticity to be recovered from \eqref{dyddpl}-\eqref{movic}. We define a pair of fields $(u, U^{(E)})$, or equivalently $(u, U^{(P)})$, satisfying \eqref{dyddpl}-\eqref{movic} as a classical linear elastic solution with stress given by $C\,(Grad \, u)^s$ provided $Curl\, U^{(E)}=0$ (or equivalently $Curl\, U^{(P)}=0$) and the total displacement $u$ satisfies the equations obtained from \eqref{dymtn}-\eqref{movic} on formally setting $U^{(P)}= 0$. 
It is now easily shown   that necessary and sufficient conditions for a solution of \eqref{dyddpl}-\eqref{movic} to be a classical elastic solution with stress given by $C\,(Grad \, u)^s$ are \eqref{qs_reduct} or \eqref{qs_reduct_SH}.
\end{rem}

To emphasise the crucial importance of the boundary condition \eqref{qschbc} when working with the Stokes Helmholtz representation,  it suffices to consider $\partial\Omega_{1}=\emptyset$ and to suppose the contrary:
\begin{equation}
\label{not}
\chi.N\neq 0,\qquad \mbox{for some 2-d neighborhood in} \ \ \partial\Omega.
\end{equation}
We show that subject to \eqref{not}, a solution of \eqref{dyddpl}-\eqref{movic}  fails to be a classical linear elastic solution with stress given by $C\, (Grad\,u)^s$  when $\alpha=0$, $r=0$, and $s=0$.




Assumption $\alpha=0$ by \eqref{qsddch}  implies
\begin{equation}
\label{gradchi}
\chi=Grad\,\Psi,
\end{equation}
where by \eqref{qsdiv}, $\Psi$  is any harmonic vector-valued function. In particular, suppose
\begin{equation}
\label{psibc}
\Psi(x)\neq 0,\qquad \mbox{for some $x$ in} \, \, \partial\Omega
\end{equation}
and
\begin{equation}
\label{normpsi}
\oint_{\partial\Omega}\frac{\partial\Psi}{\partial N}\,dS=0.
\end{equation}
Then, for non-constant $\Psi$, \eqref{not} is satisfied and consequently $\partial\Psi/\partial n\neq 0$ on $\partial\Omega$. We further require that $\chi$ is not a skew-symmetric tensor field (in which case it would be a constant  by the argument leading to \eqref{rig}). This can always be achieved by considering $\Psi$ subject to a  suitable non-vanishing Dirichlet boundary condition. As an explicit example, consider $\Psi_i = a_{ij}x_j$ in $\Omega$, where $a_{ij}$ is a constant invertible symmetric matrix and both \eqref{gradchi} and \eqref{not} are  satisfied.

Furthermore, since $r=s=0$, \eqref{zeq} and \eqref{zbc} imply $z=c_{1}t+c_{2}$ for constants $c_{1},\,c_{2}$, and in consequence we have
\begin{equation}
\label{upgradpsi}
U^{(P)}(x,t)=-\chi(x,t)=-Grad\,\Psi.
\end{equation}

Subject to conditions \eqref{conv} on the elastic modulus tensor $C$, uniqueness theorems in linear elastostatics state that the condition \eqref{qs_reduct} 
for $(u, U^{(P)})$  to be a classical linear elastic solution is satisfied if and only if $\Psi(x)$ is a rigid body displacement. For the $U^{(P)}$ given by
\eqref{upgradpsi},  construct a solution $u$ to \eqref{dyddpl}-\eqref{movic}; clearly this pair $(u, U^{(P)})$ is not a classical linear elastic solution even though it satisfies $\alpha = 0$, $r = 0$, and $s = 0$ and \eqref{qsddch} and \eqref{qsdiv} hold. On the other hand, under these conditions  and  the boundary condition $\chi. N = 0$, we conclude that $U^{(P)} = 0$ on $\Omega$. Consequently, a pair $(u, U^{(P)})$ that is a solution to \eqref{dyddpl}-\eqref{movic} with this $U^{(P)}$ is a classical linear elastic solution.
We have demonstrated the significance of including boundary condition \eqref{qschbc}.

\section{Example: Single screw dislocation moving in the whole space}\label{scdisl}

In this section we construct dynamic solutions for the motion of a single screw dislocation represented by an identical dislocation density field but with markedly different stress fields, in sharp contrast with the quasi-static result for the same case with traction free boundary conditions. The fundamental reason for the difference is as follows: as explained in Sec. \ref{plast-Volterra-Love} (and with the notational agreement at the beginning of Sec.\ \ref{qsbvp} of referring to $u^d$ as $u$), the stress in the quasi-static case is given by $C(Grad\,u_1 +\chi)^s$, with $u_1 = u -z$ and $Div\, C (Grad\,u_1 + \chi)^s = 0$, so that if $\chi$ does not change, the stress does not as well, all changes of $z$ being `absorbed' by $u$. In contrast, in the dynamic case, the stress still is given by $C(Grad\,u_1 +\chi)$, with $u_1 = u-z$, but the governing equation for $u_1$ now is affected by the general time-dependence of $z$:
\[
\rho \ddot{u}_1 = Div \,C (Grad\,u_1 + \chi)^s - \rho \ddot{z}.
\]
Hence, it becomes clear that an infinite collection of stress fields can be associated with the motion of a single dislocation (of any type) if the only constraints on the stress field are the balance of linear momentum and the specification of the dislocation by its Burgers vector and the location of its line (alternatively, the dislocation density), this being arranged by arbitrary, time-dependent gradient parts in $U^{(P)}$.

We assume  that the whole space is occupied by a linear homogeneous isotropic compressible elastic material,  in which  a single screw dislocation moves with uniform  speed $v$ along the positive $x_{1}$-axis in the slip plane perpendicular to the $x_{2}$-axis. The prescribed time-dependent dislocation density is given by
\begin{equation}
\label{dddy}
\alpha(x_{1},x_{2},t)=|b|\delta(x_{1}-vt)\delta(x_{2})e_{3}\otimes e_{3},
\end{equation}
where, as before,  $e_{i},\,i=1,2,3$ are the unit coordinate vectors,  $\delta(.)$ is the Dirac delta function, and $|b|$ is the constant magnitude of the Burgers vector. In accordance with the discussion of  Section~\ref{mov}, the plastic distortion tensor is completely represented by the Stokes-Helmholtz decomposition  \eqref{dysh}. For  $ (x,t)\in \RR^{3}\times [0,T)$, the incompatible tensor potential function $\chi(x,t)$ satisfies the system
\begin{eqnarray}
\label{cuchdddy}
Curl\,\chi &=& \alpha,\\
\label{cuchdy}
&=& |b|\delta(x_{1}-vt)\delta(x_{2})e_{3}\otimes e_{3},\\
\label{divchdy}
Div\,\chi&=& 0,
\end{eqnarray}
and the vector potential $z(x,t)$ satisfies
\begin{equation}
\label{zdy}
Div\,Grad\,\dot{z}=r, \qquad (x,t)\in \RR^{3}\times [0,T),
\end{equation}
for appropriately chosen  $r(x,t)$. Two separate choices $r^{(1)},\,r^{(2)}$ are discussed.

\begin{rem}
It follows from (\ref{cuchdy}) and (\ref{divchdy}) that without loss we may assume $\chi(x,t)$ is independent of the space variable $x_{3}$. Likewise, when $r(x,t)=r(x_{1},x_{2},t)$ there is no loss in supposing that $z(x,t)$ is independent of $x_{3}$.
\end{rem}

\begin{rem}
Because the problem is considered on the whole space, we set $s=0$ and replace   boundary conditions by the requirement that both $\chi$ and $z$  asymptotically vanish to sufficient order as $x_{\beta}x_{\beta}\rightarrow \infty,\,\beta =1,2$. Certain initial conditions, however, are still needed and are introduced as required.
\end{rem}

\subsection{Determination of $\chi$}\label{dynchi}
As  with the example of Section~\ref{stsc},  we do not employ \eqref{cuchdddy}-\eqref{divchdy}  to determine the tensor potential $\chi(x,t)$. Instead, we employ the tensor potential field analogous to $A$ introduced   in \eqref{achi}, to write
\begin{equation}
\label{chiArelmov}
\chi=Curl\,A,\qquad Div\,A=0,\qquad (x,t)\in \RR^{2}\times [0,T).
\end{equation}
 The time variable enters into the determination of both $A$ and $\chi$  as a parameter.

The corresponding Poisson equation becomes
\begin{equation}
\label{LaAdy}
\Delta\,A=-\alpha =-|b| \delta(x_{1}-vt)\delta(x_{2})e_{3}\otimes e_{3},\qquad (x,t)\in\RR^{2}\times [0,T),
\end{equation}
whose solution, by analogy with \eqref{Aesoln}, consists of the non-trivial component
\begin{equation}
\label{Adysol}
A_{33}(x_{1},x_{2},t) =-\frac{|b|}{2\pi}\ln{\bar{R}},\qquad (x,t)\in(\RR^{2}\times [0,T),
\end{equation}
where
\begin{equation}
\label{bardef}
\bar{R}^{2}=(x_{1}-vt)^{2}+x_{2}^{2}.
\end{equation}

It follows from Liouville's Theorem that all other components of $A$ vanish subject to appropriate asymptotic behaviour.

Insertion of \eqref{Adysol} into \eqref{chiArelmov}  establishes that the non-zero components of $\chi$ become
\begin{eqnarray}
\label{chi31movex}
\chi_{31}&=& A_{33,2}=-\frac{|b|}{2\pi} \frac{x_{2}}{\bar{R}^{2}},\\
\label{chi32movex}
\chi_{32}&=& -A_{33,1}=\frac{|b|}{2\pi}\frac{(x_{1}-vt)}{\bar{R}^{2}}.
\end{eqnarray}

\subsection{Determination of $z$ for given $r(x,t)=r^{(1)}(x_{1},x_{2},t)$:  First choice}\label{dyz}
Recall that  equations are specified for the vector $\dot{z}$ and not $z$, which must subsequently be found by a time integration.

The vector $r^{(1)}(x_{1},x_{2},t)$ must satisfy the compatibility relation \eqref{movrscomp} on $\RR^{2}$ for each $t$. Accordingly, we select $r^{(1)}$  to have the trivial  components   $r^{(1)}_{\gamma}=0,\,\gamma=1,2$. Therefore, the corresponding components  $\dot{z}_{\gamma}$ are  harmonic in the whole space and in view of the assumed spatial asymptotic behaviour,  $\dot{z}_{\gamma}$  vanishes  by Liouville's Theorem. Consequently,
\begin{equation}
\label{initzgam}
z_{\gamma}(x_{1},x_{2},t)=z^{(0)}_{\gamma}(x_{1},x_{2}),\qquad (x,t)\in \RR^{2}\times [0,T).
\end{equation}

Next, we choose
\begin{equation}
\label{rdy}
r^{(1)}_{3}(x_{1},x_{2},t)=|b|v\delta(x_{1}-vt)\delta^{\prime}(x_{2}),
\end{equation}
so that \eqref{movrscomp} is satisfied, and obtain
\begin{equation}
\label{z3dot}
\dot{z}_{3,\beta\beta}=|b|v\delta(x_{1}-vt)\delta^{\prime}(x_{2}),\qquad (x,t)\in \RR^{2}\times [0,T).
\end{equation}
The last equation may be integrated directly or alternatively, we have    from \eqref{cuchdddy} -\eqref{divchdy} that
\begin{equation}
\nonumber
\Delta\,\chi =-Curl\,\alpha.
\end{equation}
In particular,
\begin{equation}
\nonumber
\chi_{31,\gamma\gamma}=-\alpha_{33,2}=-|b|\delta(x_{1}-vt)\delta^{\prime}(x_{2}).
\end{equation}
which is the same  as equation \eqref{z3dot} apart from the multiplicative constant $-v$. Consequently, in view of \eqref{chi31movex}, we obtain
\begin{equation}
\label{zdy}
\dot{z}_{3}(x_{1},x_{2},t)=\frac{|b|vx_{2}}{2\pi \bar{R}^{2}}.
\end{equation}
 whose integration with respect to time  yields
\begin{equation}
\label{inzdy}
z_{3}(x_{1},x_{2},t)=-\frac{|b|}{2\pi}\left[\tan^{-1}\frac{(x_{1}-vt)}{x_{2}}-\tan^{-1}\frac{x_{1}}{x_{2}}\right]+z^{(0)}_{3}(x_{1},x_{2}),
\end{equation}
where the range of the function $\tan^{-1}$ is assumed to be $\left[ - \frac{\pi}{2}, \frac{\pi}{2} \right]$.
It remains to calculate  the initial vector $z^{(0)}(x_{1},x_{2})$.

\subsubsection{Initial values and solution for $z$}\label{init}
On denoting initial values of quantities  by a superposed zero, we have that the initial stress in terms of the initial total displacement and plastic distortion becomes
\begin{eqnarray}
\nonumber
\sigma^{(0)}_{ij}(x_{1},x_{2})&=& \lambda\left(u^{(0)}_{k,k}-U^{(P)(0)}_{kk}\right)\delta_{ij}+\mu\left(u^{(0)}_{i,j}+u^{(0)}_{j,i}-U^{(P)(0)}_{ij}-U^{(P)(0)}_{ji}\right)\\
\nonumber
&=& \lambda u^{(0)}_{\beta,\beta}\delta_{ij}+\mu\left(u^{(0)}_{i,j}+u^{(0)}_{j,i}\right)-\lambda\left(z^{(0)}_{k,k}-\chi^{(0)}_{kk}\right)\delta_{ij}\\
\nonumber
& & -\mu\left(z^{(0)}_{i,j}+z^{(0)}_{j,i}\right)+\mu\left(\chi^{(0)}_{ij}+\chi^{(0)}_{ji}\right),
\end{eqnarray}
where  the Stokes-Helmholtz decomposition of $U^{(P)(0)}(x)$ (cp., (\ref{insh})) is
\begin{equation}
\label{initscrsh}
U^{(P)(0)}(x)=Grad\, z^{(0)}(x)-\chi^{(0)}(x),\qquad x\in\Omega.
\end{equation}


In preparation for the description of initial data, we recall that  $H(.)$ and $\mbox{sign$(.)$}$  denote the \textit{Heaviside}  and \textit{sign}  generalised functions respectively.  We also set
\begin{align}
\label{notone}
\omega^{2}&= 1-\frac{v^{2}}{c^{2}},\qquad c^{2}=\frac{\mu}{\rho},\\
\label{nottwo}
 B^{2}&=x^{2}_{1}+\omega^{2} x^{2}_{2},\quad D^{2}=x^{2}_{1}+\omega x^{2}_{2},
\end{align}
 and assume that
\begin{equation}
\label{ompos}
\omega >0.
\end{equation}

The initial data  (\ref{alphainit}) and (\ref{kininit}) for $x\in \RR^{2}$  is specified to be
\begin{align}
\label{idd}
\alpha^{(0)}(x)&= \lim_{t\rightarrow 0}{\alpha(x,t)}=|b|\delta(x_{1})\delta(x_{2})\textbf{e}_{3}\otimes\textbf{e}_{3},\\
\label{itra}
l_{\gamma}(x) &= f_{\gamma}(x) = m_{\gamma}(x)=0,\qquad \gamma=1,2,\\
\label{inl}
 l_3(x)  & = - \frac{|b|}{2 \pi} \left[ \tan^{-1} \left( \frac{x_1}{\omega x_2} \right) + \frac{\pi}{2} sign ( x_2) \right]\\
\label{ivel}
f_{3}(x) &= \frac{|b|}{2\pi}\frac{v\omega x_{2}}{B^{2}}\\
\label{iacc}
m_{3}(x) & = c^2 \left[ l_{3,\beta \beta} + |b| H(x_1) \delta'(x_2) \right].
\end{align}
The choice \eqref{idd} is natural; \eqref{itra}-\eqref{ivel} are motivated by classical solutions of a uniformly moving screw dislocation \cite{hl82}. The choice \eqref{iacc} is motivated by \eqref{z_init_gov} and \eqref{crucial}.

The calculation of $\chi^{(0)}(x)$ is similar to  that   in Section~\ref{mov} and not only shows that $\chi^{(0)}(x)$ is independent of $x_{3}$ but also in particular that
\begin{eqnarray}
\label{idiv}
\chi^{(0)}_{ij,j}&=&0,\\
\nonumber
\chi^{(0)}_{31}(x_{\beta})&\neq& 0,\qquad \chi^{(0)}_{32}(x_{\beta})\neq 0,\qquad \beta=1,2,
\end{eqnarray}
while all other components of $\chi^{(0)}$ vanish identically.  Although explicit expressions are  not required, it is useful to note the relations
\begin{equation}
\label{chispec}
\chi^{(0)}_{kk}(x_{1},x_{2})=0, \qquad \chi^{(0)}_{ji,j}(x_{1},x_{2})=0.
\end{equation}


We suppose  there is sufficient continuity for the equations of motion to be valid in the limit as $t\rightarrow 0^{+}$. In consequence, we have
\begin{equation}
\label{geninmtn}
(\lambda+\mu)l_{k,k i}+\mu l_{i,kk}-(\lambda+\mu)z^{(0)}_{k,ki}-\mu z^{(0)}_{i,kk}
 +\lambda \chi^{(0)}_{kk,i}+\mu\left(\chi^{(0)}_{ij,j}+\chi^{(0)}_{ji,j}\right)=\rho m_{i}.
\end{equation}

Initial data  (\ref{itra}) and (\ref{inl}) imply that
\begin{equation}
\nonumber
l_{k,ki}=l_{\beta,\beta i}=0,\qquad l_{i,kk}=l_{i,\beta\beta}=l_{3,\beta\beta},
\end{equation}
and in conjunction with (\ref{idiv}) and  (\ref{chispec}) reduce (\ref{geninmtn}) to
\begin{equation}
\label{inmtn}
\mu l_{i,\beta\beta}(x_{1},x_{2})-(\lambda+\mu)z^{(0)}_{k,ki}(x)-\mu z^{(0)}_{i,kk}(x)=\rho m_{i}(x_{1},x_{2}).
\end{equation}

On setting $v_{i}(x)=z^{(0)}_{i,3}$,  after differentiation of (\ref{inmtn}) with respect to $x_{3}$ we obtain the equation
\begin{equation}
\nonumber
(\lambda+\mu)v_{k,ki}+\mu v_{i,kk}=0, \qquad x\in\mathbb{R}^{3}.
\end{equation}
Assume that $v$ vanishes asymptotically at large spatial distances. Then, Liouville Theorem yields $v_{i}(x) =0$ and we conclude that $z^{(0)}_{i}(x)$ is independent of $x_{3}$.

Consequently,  system  (\ref{inmtn}) may be regarded as the linear elastic equilibrium equations
for $z^{(0)}(x_{1},x_{2})$  subject to a pseudo-body force independent of $x_{3}$. 
When $i=\gamma=1,2$,  (\ref{inmtn}) and (\ref{itra}) yield the plane elastic  system
\begin{equation}
\nonumber
(\lambda+\mu)z^{(0)}_{\beta,\beta\gamma}(x_{1},x_{2})+\mu z^{(0)}_{\gamma,\beta\beta}(x_{1},x_{2})=0,\qquad x\in\RR^{2},
\end{equation}
which by  Liouville's Theorem leads to $z^{(0)}_{\gamma}(x_{1},x_{2})=0$.

When $i=3$, (\ref{inmtn})  becomes
\begin{equation}
\label{z_init_gov}
l_{3,\beta\beta}-z^{(0)}_{3,\beta\beta} = c^{-2} m_{3},
\end{equation}
which due to (\ref{inl}) and (\ref{iacc}) has a solution
\begin{equation}\label{z_init_sol}
z^{(0)}_{3}(x_{\beta})=-\frac{|b|}{4}sign(x_{2})-\frac{|b|}{2\pi}\tan^{-1} \left(\frac{x_{1}}{x_{2}}\right),
\end{equation}
which is verified as follows.

The initial value of $z_3$ has a discontinuity of $-|b|$ across the positive $x_1$ axis. We assume that $\tan^{-1} \left( \frac{x_1}{x_2} \right)$ takes the value of $0$ on the $x_1$-axis and note that $sign(x_2) = -1$ on the $x_1$-axis as well as the fact that $\tan^{-1} \left( \frac{x_1}{x_2} \right)$ has a jump of $sign(x_1) \pi$ across the $x_1$-axis which affects derivatives w.r.t $x_2$. Hence, denoting $R^2 = x_1^2 + x_2^2$, we have
\begin{align}\label{z3_deriv}
&z_{3,1}^{(0)}  = - \frac{|b|}{2 \pi} \frac{x_2}{R^2}; \qquad z_{3,2}^{(0)}  = \frac{|b|}{2 \pi} \frac{x_1}{R^2} - \frac{|b|}{2} sign(x_1) \delta(x_2) - \frac{|b|}{2} \delta(x_2) = \frac{|b|}{2 \pi} \frac{x_1}{R^2} - |b| H(x_1) \delta(x_2)\notag\\
& \frac{|b|}{\pi} \frac{x_2 x_1}{R^4}  = z_{3,11}^{(0)} ; \qquad \frac{|b|}{\pi} \frac{x_2 x_1}{R^4}  = - z_{3,22}^{(0)}  - |b| H(x_1) \delta'(x_2) \notag\\
& \implies z^{(0)}_{3,\beta \beta} =  - |b| H(x_1) \delta'(x_2).
\end{align}
This completes the derivation of initial values of the vector $z$.\footnote{ Equations \eqref{z3_deriv}$_2$ have two interpretations. In one, a distribution denoted by $\frac{|b|}{\pi} \frac{x_2 x_1}{R^4}$ is defined in two different ways. Using \eqref{z_init_sol} it can be checked that both definitions indeed define the same distribution and therefore \eqref{z3_deriv}$_3$ follows in the sense of distributions. Alternatively, the two statements in \eqref{z3_deriv}$_2$ may be interpreted as equalities involving functions with non-integrable singularities some of whose cancellation again results in \eqref{z3_deriv}$_3$.} 


Substitution of \eqref{z_init_sol} in \eqref{inzdy}  gives for $(x,t)\in\RR^{2}\times [0,T)$
\begin{eqnarray}
\label{74z}
z_{\gamma}(x_{1},x_{2},t)&=& 0, \nonumber\\
z_{3}(x_{1},x_{2},t)&=&  -\frac{|b|}{2\pi}\tan^{-1}\left(\frac{x_{1}-vt}{x_{2}}\right) - \frac{|b|}{4}sign(x_{2}).
\end{eqnarray} 
\subsection{Plastic distortion, total displacement, elastic distortion and stress}
 Expressions for $\chi$ and $z$ when substituted in (\ref{dysh}) verify that all components of $U^{(P)}(x_{1},x_{2},t)$ vanish identically on $\RR^{2}\times [0,T)$ apart from $U^{(P)}_{32}$. In particular, we have
\begin{eqnarray}
\nonumber
U^{(P)}_{31}&=& z_{3,1}-\chi_{31}\\
\nonumber
&=& -\frac{|b|}{2\pi}\frac{x_{2}}{\bar{R}^{2}}+\frac{|b| x_{2}}{2\pi\bar{R}^{2}}\\
\label{pldy31}
&=& 0,\\
\nonumber
U^{(P)}_{32}&=& z_{3,2}-\chi_{32}\\
\nonumber
&=&- |b|H(x_{1}-vt)\delta(x_{2})+\frac{|b|(x_{1}-vt)}{2\pi\bar{R}^{2}}-\frac{|b|(x_{1}-vt)}{2\pi\bar{R}^{2}}\\
\label{pldy32}
&=&- |b|H(x_{1}-vt)\delta(x_{2}),
\end{eqnarray}
where (\ref{pldy32}) is well-known in the literature. (See,e.g., \cite{hl82}.)  The formulae use similar manipulations as used in deriving \eqref{z3_deriv}.

An expression for the total displacement $u(x,t)$ is derived from the equations of motion and initial conditions. The whole space  is occupied by a linear isotropic homogeneous compressible elastic body, for which  the requisite equations, given by
\begin{equation}
\label{eqnmtn}
(\lambda+\mu)u_{j,ji}+\mu u_{i,jj}-\mu\left(U^{(P)}_{ij}+U^{(P)}_{ji}\right)_{,j}=\rho\ddot{u}_{i},\qquad (x,t)\in \RR^{3}\times [0,T),
\end{equation}
 correspond to the linear elastodynamics  equations of motion with time-varying body-force. The solution may be found using the spatial-temporal elastic  Green's function for the three-dimensional whole space. (See,e.g., \cite{kupradze1963dynamical}.) We prefer, however, to employ
properties established in Sections~\ref{dynchi} and ~\ref{init} for the potential functions $\chi(x_{1},x_{2},t)$ and $z(x_{1},x_{2},t)$ appearing in  the Stokes-Helmholtz  decomposition  (\ref{dysh}) for $U^{(P)}$. In particular,  $\chi_{31},\chi_{32}$  are the only non-zero components   so that $Div\,\chi^{s}=0$ since  $Div\,\chi$ vanishes. Moreover, the components of the vector $z$ are independent of $x_{3}$, while $z_{1}=z_{2}=0$ and therefore the nonzero components of $Grad\,z$ are $z_{3,\beta}$. In consequence, (\ref{eqnmtn}) may be written as
\begin{equation}
\label{sceqn}
(\lambda+\mu)u_{j,ji}+\mu u_{i,jj}-\mu z_{i,\beta\beta}=\rho\ddot{u}_{i},\qquad (x,t)\in\RR^{3}\times [0,T).
\end{equation}

Thus, the total displacement  is explicitly independent of $\chi$. The dislocation density $\alpha(x,t)$  given by \eqref{dddy} , however,  is implicitly present because the particular density representing the uniformly moving screw dislocation determines the form  of $\chi$ which results in its  absence from (\ref{sceqn}).

On setting $V_{i}(x,t)=u_{i,3}(x,t)$, we obtain from (\ref{sceqn}) the further reduction
\begin{equation}
\label{vmtn}
(\lambda+\mu)V_{j,ji}+\mu V_{i,jj}=\rho \ddot{V}_{i}, \qquad (x,t)\in \mathrm{I\!R\!}^{3}\times [0,T),
\end{equation}
to which are adjoined the homogeneous initial conditions $V_{i}(x,0)=l_{i,3}(x_{\beta})=0$ and $\dot{V}_{i}(x,0)=f_{i,3}(x_{\beta})=0$. Uniqueness theorems in linear elastodynamics   combined with the assumed spatial asymptotic behaviour imply that $V(x,t)$ identically vanishes. In consequence, $u(x,t)$ is independent of $x_{3}$.

When  $i= 1,2$,  the equations of motion  (\ref{sceqn})  reduce to
\begin{equation}
\nonumber
(\lambda+\mu)u_{\beta,\beta\gamma}+\mu u_{\gamma,\beta\beta}=\rho \ddot{u}_{\gamma}.\qquad (x,t)\in\mathrm{I\!R\!}^{2}\times [0,T),
\end{equation}
subject to homogeneous initial data $l_{\gamma}=f_{\gamma}=0$. An appeal to the linear elastodynamic uniqueness theorem in two dimensions shows  that $u_{\gamma}(x_{\beta},t)$ is identically zero.

Next, set $i=3$ and recall that $u$ and $z$ are independent of $x_{3}$ so that  (\ref{sceqn}) becomes
\begin{equation}
\label{totu}
u_{3,\beta\beta}-z_{3,\beta\beta}=c^{-2}\ddot{u}_{3},\qquad c^{2}= \frac{\mu}{\rho}.
\end{equation}
Equation \eqref{74z} implies
\begin{equation}
\label{crucial}
z_{3,\beta\beta}(x_{1},x_{2},t)=-|b|H(x_{1}-vt)\delta^{\prime}(x_{2}).
\end{equation}

 We  solve (\ref{totu}) by means of   the Lorentz transformation given by
\begin{equation}
\label{lxm}
\zeta_{1}=\frac{(x_{1}-vt)}{\omega},\qquad \zeta_{2}=x_{2}, \qquad \tau=\frac{(t-vc^{-2}x_{1})}{\omega},
\end{equation}
where $\omega^{2}=1-v^{2}/c^{2}$ and it is assumed that $\omega>0$.  In terms of this coordinate transformation, (\ref{totu}) becomes
\begin{equation}
\nonumber
\frac{\partial^{2} u_{3}}{\partial\zeta_{\beta}\partial\zeta_{\beta}}+|b|H(\zeta_{1})\delta^{\prime}(\zeta_{2})=c^{-2}\frac{\partial^{2}u_{3}}{\partial\tau\partial\tau},
\end{equation}
where the relation $H(ax)=H(x), a>0$ is used, and $u_{3}$ is regarded as a function of $\zeta_{\beta}$ and $\tau$. Under the customary (self-consistent) ansatz that $u_3(\zeta, \tau)$ is independent of $\tau$(see \eqref{zaltsol} below), we have 

\begin{equation}
\label{dystat}
\frac{\partial^{2} u_{3}}{\partial\zeta_{\beta}\partial\zeta_{\beta}}+|b|H(\zeta_{1})\delta^{\prime}(\zeta_{2})=0.
\end{equation}

Put
\begin{equation}
\label{tilr}
\widetilde{r}^{2}=(\zeta_{\beta}-\xi_{\beta})(\zeta_{\beta}-\xi_{\beta}).
\end{equation}
Use of the spatial Green's function successively gives
\begin{eqnarray}
\nonumber
u_{3}(\zeta_{1},\zeta_{2},\tau)&=& -\frac{|b|}{2\pi}\int_{\RR^{2}}H(\xi_{1})\delta^{\prime}(\xi_{2})\ln \widetilde{r}\,d\xi_{1}d\xi_{2}\\
\nonumber
&=&\frac{|b|}{2\pi}\int_{\RR^{2}}\frac{\partial}{\partial\xi_{2}}\left(\ln\widetilde{r}\right)H(\xi_{1})\delta(\xi_{2})\,d\xi_{1}d\xi_{2}\\
\nonumber
&=&-\frac{|b|\zeta_{2}}{2\pi}\int^{\infty}_{-\infty}\frac{H(\xi_{1})}{(\zeta_{1}-\xi_{1})^{2}+\zeta_{2}^{2}}\,d\xi_{1}\\
\nonumber
&=&-\frac{|b|\zeta_{2}}{2\pi}\int^{\infty}_{0}\frac{1}{(\zeta_{1}-\xi_{1})^{2}+\zeta_{2}^{2}}\,d\xi_{1}\\
\nonumber
&=& \frac{|b|}{2\pi}\tan^{-1}\left(\frac{\zeta_{1}-\xi_{1}}{\zeta_{2}}\right)\mid^{\xi_{1}=\infty}_{\xi_{1}=0}\\
\label{z3soln}
&=& -\frac{|b|}{2\pi}\left[\tan^{-1}\left(\frac{\zeta_{1}}{\zeta_{2}}\right)-\frac{\pi}{2}S(\zeta_{1},\zeta_{2})\right].
\end{eqnarray}
The generalised function $S(\zeta_{1},\zeta_{2})$, defined by
\begin{equation}
\label{Sdef}
\frac{\pi}{2}S(\zeta_{1},\zeta_{2})=\lim_{\xi_{1}\rightarrow\infty}{\tan^{-1}\left(\frac{(\zeta_{1}-\xi_{1})}{\zeta_{2}}\right)},
\end{equation}
  upon evaluation   reduces to 
\begin{equation}
\label{sgns}
S(\zeta_{1},\zeta_{2}) 
= -sign(\zeta_{2}).
\end{equation}

Insertion of \eqref{sgns} into  (\ref{z3soln}) leads to  the representation
\begin{equation}
\label{zaltsol}
u_{3}(\zeta_{1},\zeta_{2},\tau)=-\frac{|b|}{2\pi}\left[\tan^{-1}\left(\frac{\zeta_{1}}{\zeta_{2}}\right)+\frac{\pi}{2} sign (\zeta_{2})\right].
\end{equation}
In terms of the  original coordinates, (\ref{zaltsol}) becomes
\begin{equation}\label{u_soln}
u_{3}(x_{1},x_{2},t) = -\frac{|b|}{2\pi}\left[\tan^{-1}\left(\frac{(x_{1}-vt)}{\omega x_{2}}\right)+\frac{\pi}{2}\mbox{sign $(x_{2})$}\right].
\end{equation}

It is  easily  verified  by direct substitution that the last expressions  for $u_{3}$ combined with $u_{\gamma}(x,t)=0$ identically satisfy the equations of motion (\ref{sceqn}), and  are compatible with   initial conditions specified in Section~\ref{init}.

Components of the elastic distortion  are derived from the identity
\begin{equation}
\nonumber
U^{(E)}=Grad\, u-U^{(P)},
\end{equation}
which after substitution from (\ref{pldy31}), (\ref{pldy32}),  together with  expressions (\ref{zaltsol}) and (\ref{u_soln}) for $u_{3}$ gives the non-zero components of $U^{(E)}$ as
\begin{eqnarray}
\nonumber
U^{(E)}_{31}&=& u_{3,1}-U^{(P)}_{31}\\
\label{eldy31}
&=&-\frac{|b|}{2\pi}\frac{\omega x_{2}}{(x_{1}-vt)^{2}+\omega^{2}x_{2}^{2}},\\
\nonumber
U^{(E)}_{32}&=& u_{3,2}-U^{(P)}_{32}\\
\nonumber
&=&-|b|H(x_{1}-vt)\delta(x_{2})+\frac{|b|}{2\pi}\frac{\omega(x_{1}-vt)}{(x_{1}-vt)^{2}+\omega^{2}x_{2}^{2}}+|b|H(x_{1}-vt)\delta(x_{2})\\
\label{eldy32}
&=& \frac{|b|}{2\pi}\frac{\omega(x_{1}-vt)}{(x_{1}-vt)^{2}+\omega^{2}x_{2}^{2}}.
\end{eqnarray}

Non-zero components of the stress, derived  from the linear constitutive relations \eqref{ssrel},  are given by
\begin{eqnarray}
\label{sdy31}
\sigma_{31}&=& \mu U^{(E)}_{31}= -\frac{\mu |b|}{2\pi}\frac{\omega x_{2}}{(x_{1}-vt)^{2}+\omega^{2}x_{2}^{2}},\\
\label{sdy32}
\sigma_{32}&=& \mu U^{(E)}_{32}=\frac{\mu |b|}{2\pi}\frac{\omega(x_{1}-vt)}{(x_{1}-vt)^{2}+\omega^{2}x_{2}^{2}}.
\end{eqnarray}
Expressions \eqref{sdy31} and \eqref{sdy32} are well-known in the literature (c.p.,\cite{hl82}), but usually are derived by entirely different methods.

The corresponding Burgers vector may be calculated from \eqref{beldis} using \eqref{eldy31} and \eqref{eldy32}. We have $b= (0,0,b_{3})$, where
\begin{equation}
\nonumber
b_{3}=\oint_{\partial\Sigma} U^{(E)}_{3\beta}(x,t).dx_{\beta}=|b|,
\end{equation}
and $\partial\Sigma$ is the circle of unit radius centred at the origin.

\begin{rem}
The screw dislocation moving with uniform velocity is shown by Pellegrini \cite{p10} to be the stationary limit of more generally moving screw dislocations. In particular, this author studies the relationship not only with a dynamic Peierls-Nabarro equation but also in the limit with Weertman's equation \cite{weertman1967uniformly} . See also the discussion in  \cite{m} and \cite{pp}.
\end{rem}

\subsection{Determination of $z$ for $r(x,t)=r^{(2)}(x,t)$ = 0: Second choice} \label{r2}
The specification of $r(x, t)$ to determine $Grad\, z$ in the Stokes-Helmholtz  decomposition (\ref{dysh})  of $U^{(P)}$ is arbitrary. As illustration,  we set $r(x,t)=r^{(2)}(x,t)=0$ in the problem just considered of the screw dislocation uniformly moving in the whole space with initial data  specified  by \eqref{itra}-\eqref{iacc}. The physical interpretation of this problem is as follows: consider the plastic distortion profile corresponding to a dislocation moving in the positive $x_1$ direction:
\begin{align*}
U^{(P)}_{32} (x_1,x_2,t) &= -|b| H(x_1-vt) \delta(x_2) \notag\\
U^{(P)}_{ij} (x_1,x_2,t) &= 0, i \neq 3, j \neq 2.
\end{align*}
Suppose we perform a Helmholtz decomposition of this plastic distortion field, take its time evolving compatible part (i.e., $Grad\,z$) satisfying \eqref{zdoteq} and \eqref{zdotbc} with $r = Div\,\dot{U}^{(P)}$, $s = \dot{U}^{(P)}. n$,  and initial condition \eqref{z_init_sol}, and consider the physical problem of calculating the fields of a dislocation moving in a body subjected to an additional plastic distortion field given by  $-Grad \, z$, say arising from other sources of plasticity, then the results derived below in this Section correspond to the fields of a superposition of these two evolving eigenstrain fields. Note that the prescribed  field $Grad\, z$ is also singular but has vanishing $curl$ in the sense of distributions and, consequently, the dislocation density is identical to that of the $U^{(P)}$  field alone.

The tensor potential $\chi(x,t) $ remains unaltered from the values (\ref{chi31movex}) and (\ref{chi32movex}),  but now $\dot{z}(x_{\beta},t)=0$, and consequently, $z(x_{\beta},t)=z^{(0)}(x_{\beta})$, where as proved in Section~\ref{init}, $z^{0}_{\gamma}=0$ and $z^{0}_{3}$ is given by \eqref{z_init_sol}. That is,
\begin{align}\label{z_choice2}
z_{\gamma}(x_{\beta},t)& = z^{(0)}_{\gamma}(x_{\beta})  =0 \notag\\
z_3(x,t)& = z^{(0)}_{3}(x) =-\frac{|b|}{4}sign(x_{2})-\frac{|b|}{2\pi}\tan^{-1} \left(\frac{x_{1}}{x_{2}}\right).
\end{align} 

In particular, we have
\begin{equation}
\label{inlap}
z_{3,\beta \beta} = z^{(0)}_{3,\beta\beta}=-|b|H(x_{1})\delta^{\prime}(x_{2}).
\end{equation}

In terms of  the previously introduced notation
\begin{align}
\label{dist}
R^{2}&= x_{1}^{2}+x^{2}_{2},\\
\label{modist}
\bar{R}^{2}&= (x_{1}-vt)^{2}+x^{2}_{2},
\end{align}
 the non-zero components of the plastic distortion tensor $U^{(P)}(x,t)= Grad\, z(x,t)-\chi(x,t)$, using \eqref{z3_deriv}, are given by
\begin{align}
U^{(P)}_{31}(x,t)&= - \frac{|b| x_2}{2 \pi} \left[ \frac{1}{R^2} - \frac{1}{{\bar{R}}^2} \right] \\
U^{(P)}_{32}(x,t) & = - |b|H(x_{1})\delta(x_{2}) +\frac{|b|}{2 \pi} \left[ \frac{x_1}{R^2} - \frac{(x_1- vt)}{{\bar{R}}^2} \right]
\end{align}
where $(x,t)\in\mathrm{I\!R\!}^{2}\times [0,T)$. We observe that the components of the plastic distortion tensor given by the last two expressions  are notably different from the corresponding expressions (\ref{pldy31}) and (\ref{pldy32}) in the problem with $r= r^{(1)}$.

With respect to the total displacement $u(x,t)$,  arguments developed in the previous sections  show that $u_{\gamma}(x,t)=0$, and that $u_{3}(x,t)$ is independent of $x_{3}$. Consequently, (\ref{sceqn})  becomes
\begin{equation}
\nonumber
u_{3,\beta\beta}-P(x_{\beta})=c^{-2}\ddot{u}_{3},
\end{equation}
where the time-independent pseudo-body-force $P(x_{\beta})$ is given by
\begin{equation}
\nonumber
P(x_{\beta})=z^{(0)}_{3,\beta\beta}(x_{\beta})=-|b|H(x_{1})\delta^{\prime}(x_{2}).
\end{equation}

Let
\begin{equation}
\nonumber
w(x_{\beta},t)=u_{3}(x_{\beta},t)-z^{(0)}_{3}(x_{\beta}),
\end{equation}
so that $w(x_{\beta},t)$ satisfies the two dimensional wave equation
\begin{equation}
\label{weqn}
w_{,\beta\beta}=c^{-2}\ddot{w},\qquad (x,t)\in\mathrm{I\!R\!}^{2}\times [0,T),
\end{equation}
subject to  a standard  ``radiation''  condition and,  from  \eqref{inl} and \eqref{ivel}, the  initial conditions
\begin{align}
\label{win}
w(x_{\beta},0)&= l_{3}(x_{\beta})-z^{(0)}_{3}(x_{\beta})=-\frac{|b|}{2\pi}\tan^{-1}{\frac{(1-\omega)x_{1}x_{2}}{D^{2}}},\\
\label{wvelin}
\dot{w}(x_{\beta},0)&= f_{3}(x_{\beta})=\frac{|b|}{2\pi}\frac{\omega v x_{2}}{B^{2}},
\end{align}
where for convenience we repeat the notation
\begin{equation}
\nonumber
B^{2}=x^{2}_{1}+\omega^{2}x^{2}_{2},\qquad D^{2}= x^{2}_{1}+\omega x^{2}_{2}.
\end{equation}

The solution to the initial value problem may be obtained  using  Green's function; see, for example,  Volterra as quoted in \cite[\S 5.9 D]{es75}, where other methods of solution also are reviewed.  We employ, however,   the method of  spherical means to obtain a solution of the form
\begin{eqnarray}
\nonumber
w(x_{\beta},t)&=&\frac{|b|v\omega}{(2\pi c)^{2}}\int_{\mathrm{I\!R\!}^{2}}\frac{\xi_{2}}{\left(t^{2}-\widehat{r}^{2}/c^{2}\right)^{1/2}\left(\xi^{2}_{1}+\omega^{2}\xi^{2}_{2}\right)}\,d\xi_{1}d\xi_{2}\\
\nonumber
& &+ \frac{|b|t}{(2\pi c)^{2}}\int_{\mathrm{I\!R\!}^{2}}\frac{1}{\left(t^{2}-\widehat{r}^{2}/c^{2}\right)^{1/2}}\tan^{-1}{\left\{\frac{(1-\omega)\xi_{1}\xi_{2}}{(\xi_{1}^{2}+\omega\xi^{2}_{2})}\right\}}\,d\xi_{1}d\xi_{2},
\end{eqnarray}
where, as before,
\begin{equation}
\nonumber
\widehat{r}^{2}=(x_{\beta}-\xi_{\beta})(x_{\beta}-\xi_{\beta}).
\end{equation}

The total displacement $u(x_{1},x_{2},t)$ therefore has known components $(0,0,w+z^{(0)}_{3})$ and enables the elastic distortion tensor to be calculated  from the relation $U^{(E)}= Grad\, u-U^{(P)}$. It in turn determines the stress tensor $\sigma =C\left(U^{(E)}\right)^{s}$, whose non-zero components consequently are given by
\begin{eqnarray}
\nonumber
\sigma_{3\gamma}&=&\mu U^{(E)}_{3\gamma}\\
\nonumber
&=&\mu(u_{3,\gamma}-z^{(0)}_{3,\gamma})+\mu \chi_{3\gamma}\\
&=& \mu w_{,\gamma}+\mu\chi_{3\gamma}.
\end{eqnarray}

The corresponding explicit expressions are
\begin{align*}
\sigma_{31}(x_{\beta},t) & =\mu w_{,1}(x_{\beta},t)-\frac{\mu |b|x_{2}}{2\pi \bar{R}^{2}},\\
\sigma_{32}(x_{\beta},t) & = \mu w_{,2}(x_{\beta},t)+\frac{\mu |b|}{2\pi}\frac{(x_{1}-vt)}{\bar{R}^{2}}.
\end{align*}

In view of  \eqref{beldis}, the  Burgers vector $b$ is given by $(0,0, |b|).$

\begin{rem}
As expected, the stress components derived for the problem corresponding to $r=r^{(2)}=0$ are markedly different to those for  the problem corresponding to  $r=r^{(1)}$. Indeed, let $z^{(\gamma)}(x,t),$ $\gamma=1,2,$ be given by (\ref{74z}) and (\ref{z_choice2}) respectively. By inspection, $z^{(1)}-z^{(2)}$ is not a rigid body motion and consequently the necessary and sufficient conditions of Lemma~\ref{unstress} are violated implying that  the difference between the respective stress distributions must be non-zero at least at one point in space-time. A prescribed  dislocation density  alone therefore is  insufficient to  uniquely  determine the stress in the  linear dynamic problem of dislocations.
\end{rem}


\begin{rem}
Consider the solution corresponding to the problem in which $r =r^{(1)}-r^{(2)}=r^{(1)}$. Each constituent problem has the same initial conditions which are therefore homogeneous for the difference solution. Accordingly, we have solved the problem for $r^{(1)}$ with homogeneous initial data.
\end{rem}
\begin{rem}
Consider the dynamic problem for a uniformly moving screw dislocation (specified by \eqref{dddy}) with $l = f = m = 0$ and $z(x,t) = z^{(0)} (x) = 0$ in all of space-time. Since $Div\, \chi = 0$ in space-time, we have that $u = 0$, $z = 0$ and $\chi$ given by \eqref{chi31movex}-\eqref{chi32movex} is a solution to the dynamic, and quasi-static, problem with the mentioned prescribed data.  Of course, the time dependence of a quasi-static displacement solution does not, in general, allow the satisfaction of the dynamic equations of motion, a situation that could arise, e.g., if $Div\,( C \chi) \neq 0$.
\end{rem} 

\section{Concluding Remarks} \label{conrem}
The paper explores various aspects of the
equilibrium and dynamical  equations 
for the stress and displacement fields in an elastic body  subject to a given, possibly evolving, dislocation density field. The  plasticity theory of dislocations selected  for this purpose   facilitates  examination  of  conditions for uniqueness of solutions  to  appropriate initial boundary value problems. The Stokes-Helmholtz decompostion of second order tensor fields is consistently employed in the discussion.

 To justify   the plasticity theory adopted,  we  investigated the relation  to  the apparently different  Volterra theory of dislocations by  considering
 in detail  
  the particular example of a stationary straight dislocation.  We established in a precise sense that the Volterra problem is the limit of a sequence of plasticity problems, where the plastic distortion tensor is considered as data.

Subsequent Sections are concerned wholly with the plasticity formulation, and assume the dislocation density to be  data. Necessary and sufficient conditions for unique solutions  are derived in the static, quasi-static, and dynamic dislocation problems.  It emerges that  in the  initial boundary value problem, uniqueness is not ensured by the dislocation density even in combination with standard initial and boundary data. To achieve uniqueness  requires the additional stipulation of the vector potential $z$   in  the Stokes-Helmholtz decomposition of the plastic distortion. These results on (non)-uniqueness are summarized in the statements of Propositions \ref{uniqueness_statics} and \ref{prop_5_1}, Lemmas \ref{unstress} and \ref{undisp}, Theorem \ref{uniq}, and Remarks \ref{rem_5_2} and \ref{rem_6_1}. Our conclusions  are consistent with   recent developments in  modelling  dislocation dynamics that use concepts of nonlinear plasticity theory related to the  stress-coupled evolution of dislocation density and plastic distortion, e.g. \cite{zhang2015single, Zhang2017, roy2006size}. Such models are in agreement with  the Mura-Kosevich \cite{ma63,kosevich1979crystal} kinematics  in which the plastic distortion rate is given by the product of the  dislocation density and the dislocation velocity. An interesting feature, however,  of the present work  is  that the plasticity formulation of dislocations  can admit  alternatives beyond the Mura-Kosevich specification.  Regardless of the specification, the stress and displacement can differ corresponding to the same  dislocation density.

 A subsidiary investigation obtained  necessary and sufficient conditions for reduction of  the dislocation theory   to  respective linear theories  of classical  elasticity.

The studies described in the paper prompt several 
questions that  await further discussion. Not least,  is further consideration of the problem treated in Section~\ref{plast-Volterra-Love},  and  the  separate exploration of  uniqueness  for the nonlinear  plasticity theory of   dislocations  that includes finite deformations and nonlinear elasticity.

\textit{Acknowledgements}.
 We thank the two anonymous referees for their valuable comments. A.A. would like to acknowledge a Visiting Professorship from the Leverhulme Trust and the hospitality of the Department of Mathematical Sciences at the University of Bath where this work was partially carried out; his work has also been supported in part by grants NSF-CMMI-1435624 and ARO W911NF-15-1-0239.

\bibliographystyle{amsalpha}\bibliography{basic_structure}

\end{document}